\documentclass[journal=jacsat,manuscript=article]{achemso}

\usepackage[version=3]{mhchem} 
\usepackage{amsmath} 

\usepackage{tikz}
\usetikzlibrary{shapes, arrows, positioning, calc}
\usepackage{comment}
\usepackage{graphicx}
\usepackage{subcaption}
\usepackage{textcomp,gensymb}
\usepackage{amsfonts}
\usepackage{amsmath}
\usepackage{xr}
\usepackage{multirow}
\usepackage{hyperref}
\SectionNumbersOn

\newcommand*{\MPTRJ}{$\text{MP}_\text{trj}$}
\newcommand*{\MPT}{$\text{MP}_\text{trj}\_$MPNICE}
\newcommand*{\OMA}{OMAT24$_a\_$MPNICE}
\newcommand*{\MAT}{Inorganic$\_$MPNICE}
\newcommand*{\UNI}{Hybrid$\_$MPNICE}
\newcommand*{\MTLO}{Hybrid$\_$MPNICE$\_$O}
\newcommand*{\MTLI}{Hybrid$\_$MPNICE$\_$I}
\newcommand*{\ORG}{Organic$\_$MPNICE}
\newcommand*{\ORGTB}{Organic$\_$MPNICE$\_$TB}
\newcommand*{\CORG}{Organic$\_$Crystals$\_$MPNICE}
\author{John L. Weber}
\affiliation{Schr\"odinger Inc., 1540 Broadway, 24$^{th}$ floor, New York, NY 10036}
\author{Rishabh D. Guha}
\affiliation{Schr\"odinger Inc., 1540 Broadway, 24$^{th}$ floor, New York, NY 10036}
\author{Garvit Agarwal}
\affiliation{Schr\"odinger Inc., 1540 Broadway, 24$^{th}$ floor, New York, NY 10036}
\author{Yujing Wei}
\affiliation{Schr\"odinger Inc., 1540 Broadway, 24$^{th}$ floor, New York, NY 10036}
\alsoaffiliation{Department of Chemistry, Columbia University, 3000 Broadway, New York, NY 10027}
\author{Aidan A. Fike}
\affiliation{Schr\"odinger Inc., 1540 Broadway, 24$^{th}$ floor, New York, NY 10036}
\author{Xiaowei Xie}
\affiliation{Schr\"odinger Inc., 101 SW Main Street, Suite 1300, Portland, OR 97204}
\author{James Stevenson}
\affiliation{Schr\"odinger Inc., 1540 Broadway, 24$^{th}$ floor, New York, NY 10036}
\author{Biswajit Santra}
\affiliation{Schr\"odinger Inc., 1540 Broadway, 24$^{th}$ floor, New York, NY 10036}
\author{Richard A. Friesner}
\affiliation{Department of Chemistry, Columbia University, 3000 Broadway, New York, NY 10027}
\author{Karl Leswing}
\affiliation{Schr\"odinger Inc., 1540 Broadway, 24$^{th}$ floor, New York, NY 10036}
\author{Mathew D. Halls}
\affiliation{Schr\"odinger Inc., 9868 Scranton Road, Suite 3200 San Diego, CA 92121}
\author{Robert Abel}
\affiliation{Schr\"odinger Inc., 1540 Broadway, 24$^{th}$ floor, New York, NY 10036}
\author{Leif D. Jacobson}
\affiliation{Schr\"odinger Inc., 101 SW Main Street, Suite 1300, Portland, OR 97204}
\email{leif.jacobson@schrodinger.com}

\title[An \textsf{achemso} demo]
  {Efficient Long-Range Machine Learning Force Fields for Liquid and Materials Properties}

\abbreviations{MPNICE}
\keywords{American Chemical Society, \LaTeX}

\begin{document}

%
%
%
%
%

\begin{abstract}
    Machine learning force fields (MLFFs) have emerged as a sophisticated tool for cost-efficient atomistic simulations approaching DFT accuracy, with recent message passing MLFFs able to cover the entire periodic table. We present an invariant message passing MLFF architecture (MPNICE) which iteratively predicts atomic partial charges, including long-range interactions, enabling the prediction of charge-dependent properties while achieving 5-20x faster inference versus models with comparable accuracy. We train direct and delta-learned MPNICE models for organic systems, and benchmark against experimental properties of liquid and solid systems. We also benchmark the energetics of finite systems, contributing a new set of torsion scans with charged species and a new set of DLPNO-CCSD(T) references for the TorsionNet500 benchmark. We additionally train and benchmark MPNICE models for bulk inorganic crystals, focusing on structural ranking and mechanical properties. Finally, we explore multi-task models for both inorganic and organic systems, which exhibit slightly decreased performance on domain-specific tasks but surprising generalization, stably predicting the gas phase structure of $\simeq500$ Pt/Ir organometallic complexes despite never training to organometallic complexes of any kind.
\end{abstract}

\section{Introduction}

Density Functional Theory (DFT) is the current workhorse of computational modeling of condensed phase systems. Parametrized functionals have achieved near chemical accuracy for select applications,\cite{mardirossian2017thirty} while retaining high enough efficiency such that \emph{ab initio} molecular dynamics (AIMD) simulations using DFT can reliably be run on tens to hundreds of atoms. For properties which require larger simulations or more than a few picoseconds to resolve, empirically parametrized interatomic potentials, also known as classical force fields, have been used for decades,\cite{jorgensen1996development,monticelli2013force} and are efficient enough to run hundreds of nanoseconds of molecular dynamics (MD) a day, at the expense of reduced accuracy and generalization. While parameterizing classical force fields for specific application domains, such as organic druglike molecules, has had its fair share of success, there are many notable areas where classical force fields are lacking, such as treating reactive systems and diverse inorganic materials. In recent decades, machine learning force fields (MLFFs), sometimes referred to as machine learning interatomic potentials (MLIPs), have increased the capacity of force fields by abstracting the functional form and dramatically increasing the number of learnable parameters. MLFFs have been shown to effectively reproduce DFT accuracy for targeted applications at a fraction of the cost of DFT, albeit on the order of 100 times more expensive than classical force fields.

High dimensional neural network potentials (HDNNPs) were the first MLFFs demonstrated to achieve high accuracy on large organic datasets\cite{Smith2017ani1, Smith2018ani1x, stevenson2019sani, smith2020ani}. Such architectures exhibit poor scaling with the number of supported elements due to the use of element-dependent Behler-Parinello type symmetry functions.\cite{Behler2007BPSF}  As such, efforts to produce ``general" models of this type were restricted to subsets of the periodic table, typically neutral organic molecules\cite{smith2020ani}. In order to distinguish different charge states, as well as include long-range electrostatics interactions, some models additionally predict fractional charges on each atom.\cite{ko2021fourth, jacobson2022transferable, anstine2024aimnet2}  Models such as these have been used to perform high-fidelity simulations of specific organic systems, including redox chemistry and electrolytes for Li-ion batteries.\cite{dajnowicz2022high,kocer2024machine}

The development of message passing networks\cite{schutt2017schnet, Unke2019physnet} has effectively solved the scaling of MLFFs with respect to the number of chemical elements by embedding the element in an atomic feature vector. There has also been rapid growth in the number and scale of public databases for materials properties calculated using DFT, including the Materials Project\cite{jain2013commentary} and the Open Catalyst Project.\cite{chanussot2021open, tran2023open,barroso2024open} The combination of flexible models and diverse datasets has enabled the construction of general-purpose MLFFs for inorganic materials, with a wide variety of architectures and models showcasing impressively high accuracy and generalization to unseen structures.\cite{deng2023chgnet,chen2022universal,batatia2023foundation,fu2025learning,barroso2024open,park2024scalable,choudhary2021atomistic,bochkarev2024graph, merchant2023scaling, neumann2024orb, zhang2024dpa} With the introduction of public benchmarks and leaderboards,\cite{riebesell2023matbench} model development has trended towards larger models, often incorporating equivariant geometric features in order to increase the data-efficiency of training at the expense of increased inference cost.\cite{loew2024universal}

While development of general models in the inorganic materials landscape has advanced significantly in the past two years, it has yet to make significant contact with general organic datasets, perhaps due to the significant differences in the levels of theory employed. The majority of inorganic datasets are trained on DFT data generated with the PBE functional, utilizing pseudopotentials and plane wave basis sets under periodic boundary conditions, whereas for organic molecules, all-electron calculations are used, often employing hybrid or long-range corrected functionals combined with an atom-centered basis set. The reasons behind these choices are fundamental, involving the inaccuracy of pure functionals for organic potential energy surfaces (PES)\cite{mardirossian2017thirty, dreuw2005single} and the problem of Hartree-Fock exchange in hybrid functionals introducing errors when treating zero-gap systems.\cite{paier2007does} As a result, however, the absolute energy scale of the training data is different between organic and inorganic datasets, leading to significant conflicts during training. Any attempt to mitigate this by fixing the reference theory to one functional necessarily sacrifices accuracy on some application areas.  One approach to circumvent such conflicts in reference data is to employ multi-task learning;\cite{jacobson2023leveraging}  a single model is simultaneously trained to produce multiple outputs and the user defines at inference time which reference value to use. This approach generally leads to models with more general internal representations, and can enhance performance.\cite{jacobson2023leveraging,zhang2023universal,TakShiMot2022}  Tests of models trained to very disparate datasets have been less prevalent; the DPA-2 model (ref. \citenum{zhang2023universal}) is one example, being trained on a large number of open source datasets including hybrid and GGA functionals. The benefits of increased generalization of hidden features are most pronounced when fine-tuning such a model,\cite{zhang2024dpa} and for zero-shot performance, accuracy is still dependent on the choice of the output head to use. In cases with minor differences between datasets, meta-learning techniques have been used with success to train a single output head to multiple datasets.\cite{allen2024learning} Another recent approach to train one output head to disparate datasets simply applies a linear shift to the total energies and forces of one level of theory.\cite{shiota2024taming}  While this approach minimizes training conflicts, it does so by modifying the underlying reference PES, which may result in decreased performance for specific properties.  A model capable of simultaneously reproducing the dynamics of multiple general classes of materials and with usable accuracy versus experiment, has remained out of reach.

Building on our prior charge-aware HDNNP (QRNN)\cite{jacobson2022transferable} we here describe a charge-aware, invariant message passing MLFF, referred to as Message Passing Network with Iterative Charge Equilibration, or MPNICE. MPNICE is designed to balance generality with inference cost, in order to enable large-scale simulations, both in atom count and time scale, of as many materials as possible. The rest of the paper is organized as follows: In section \ref{sec:MPNICE} we provide a brief outline of the MPNICE architecture. The results (section \ref{sec:results}) are split into three sections according to the class of material. In section \ref{sec:organic} we describe a set of MPNICE models trained for organic molecules and their ions, including models targeting molecular dynamics simulations of organic liquids, fine-scale relative accuracy on conformations, and molecular organic crystal structure prediction (CSP). We finish section \ref{sec:organic} by highlighting the ability of MPNICE to learn ionization energies of small molecules. In section \ref{sec:inorganic}, we present a set of MPNICE models trained to inorganic crystals, with benchmarks on the ranking of different structures as well as mechanical properties such as bulk and shear moduli and phonon band structures. We additionally describe how to calculate electric response properties within the MPNICE framework, demonstrating this by calculating the long-range non-analytic correction to the NaCl phonon band structure. We finish section \ref{sec:inorganic} by highlighting the zero-shot prediction of the Li diffusion barrier and thermal expansion of amorphous LiAlO$_2$ via large scale MD simulations, in excellent agreement with experiment. In section \ref{sec:hybrid} we describe hybrid models trained to both inorganic and organic datasets, including a novel method to train a hybrid model with one output head by simply restricting energy training to one dataset, relying on forces to obtain an accurate PES for the other. We present benchmarks from the previous sections, and finish with a highlight of increased generalization by optimizing a set of Pt and Ir centered organometallic complexes. In section \ref{sec:profiling}, we evaluate the performance of MPNICE as implemented in the Desmond MD engine\cite{release20191}, comparing against comparable open source models. In section \ref{sec:conclusion} we conclude, providing guidelines for the use of MPNICE pretrained models.

\section{The MPNICE architecture}\label{sec:MPNICE}

\begin{figure}[!htb]
    \centering
    \includegraphics[width=\linewidth]{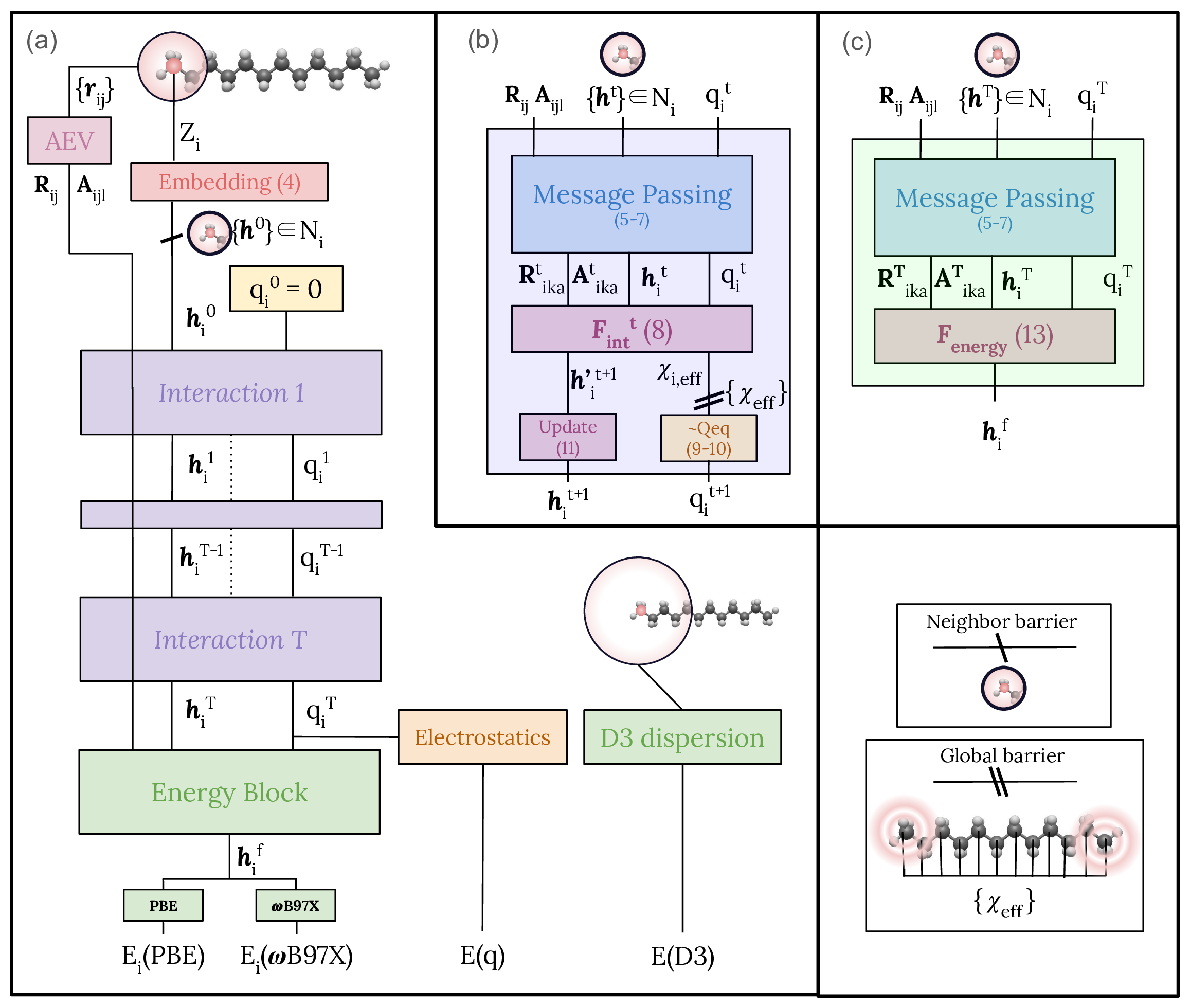}
    \caption{(a) An overview of the MPNICE architecture. Atomic features are initialized with an embedding of the chemical element, whereas geometric environments are represented using a set of atomic environment vectors (AEVs). Atomic features, atomic charges (initialized to zero), and AEVs are used as input to a set of iterative interaction blocks (b), within which messages are formed from the atomic features and AEVs, from which partial charges are predicted and atomic features are updated. After T interaction blocks an energy block (c) is used to predict a final feature vector for each atom $i$, which is used as input to a single layer MLP that predicts atomic contributions to the energy for a given level of theory. A single dash is used to denote operations which require synchronization across the nearest neighbors of the central atom, while two dashes are used to denote synchronization across the system, i.e. for charge equilibration.  Equation numbers defining the mathematical variables are given in parenthesis.}
    \label{fig:MPNICE_overview}
\end{figure}

An overview of the MPNICE architecture can be seen in Figure \ref{fig:MPNICE_overview}, panel (a). MPNICE functions as a local message passing interatomic potential, decomposing the total energy into contributions from each atom, produced as an iterative function of their local environment within a radial cutoff $r_{max}$. After every message passing iteration, atomic partial charges are predicted via an approximation to the Qeq charge equilibration algorithm\cite{jacobson2022transferable} and used as input to the next block. After $t$ interaction blocks, a final round of message passing is performed before a final multilayer perceptron (MLP) predicts the atomic energy. The last layer of the readout MLP, referred to as the output head, is generally specific to the level of theory being trained, with support for multiple output heads depending on the dataset.

The environment surrounding each central atom is represented as a set of two- and three-body Behler-Parinello symmetry functions,\cite{Behler2007BPSF} commonly known as the atomic environment vectors (AEVs).\cite{Smith2017ani1} The two-body contribution to the AEV, or radial AEV $(R(r_{ij}))$, expands the interatomic distance between the central atom $i$ and a neighboring atom $j$ using a set of evenly spaced Gaussians,
\begin{equation}
    R_{aij} = 0.25 \times \exp \left[-\eta (r_{ij} - r_a)^2\right] \times f_c(r_{ij}) 
\end{equation}
where $r_a$ denotes the radial shift of the particular Gaussian. $f_c(r_{ij})$ is a damping function which smoothly goes to zero at $r_{max}$ in order to maintain continuous parametrization as neighbors cross $r_{max}$,
\begin{equation}
   f_c(r_{ij}) = 
     \begin{cases}
         0.5 \times \cos(\frac{r_{ij}}{r_{max}} \pi) + 0.5 & r_{ij} \leq r_{max} \\
         0 & r_{ij} > r_{max} .
     \end{cases}
\end{equation}
The three-body (angular) AEV, $A_{aijl}$, between the central atom $i$ and two neighbors, $j$ and $l$, is represented by a series of basis functions,
\begin{equation}
    A_{aijl} = 2^{1-\zeta}  (1 + \cos(\theta_{ijl} - \theta_a))^{\zeta} \times \exp \left[  -\eta \left( \frac{r_{ij} + r_{il}}{2} - r_a \right)^2 \right] f_c(r_{ij}) f_c(r_{il}) .
\end{equation}

The chemical element for each atom is first encoded in a vector $\textbf{h}_i$ of arbitrary length $N_h$, which we refer to as the node feature vector (NFV). This is initialized by a learned linear mixing of a one-hot embedding, $\Omega$, of the chemical element, $Z_i$,
\begin{equation}
    h_{ik}^{(0)} = \sum_b W^h_{kb} \times \Omega[Z_i]_b + B_k.
\end{equation}
Messages are then constructed using the NFVs of the neighboring atoms and the respective AEVs, summing over all neighbors within a defined cutoff. To reduce the size of the messages, we inject a linear transformation of the pairwise node features for both the radial and angular messages, $W^r_{\tilde{k}k}$ and $W^a_{\tilde{k}k}$, to a smaller dimension of message features, typically 32, 
\begin{equation}
    \tilde{R}^t_{i\tilde{k}a} = \sum_j^{N(i)} \left[\sum_k W^{r}_{\tilde{k}k}h^{t}_{ik} h^{t}_{jk}\right] R_{aij} ,
\end{equation}
\begin{equation}
    \tilde{A}^t_{i\tilde{k}a} = \sum_l^{\tilde{N}(i)} \sum_{j \neq l}^{\tilde{N}(i)} \left[\sum_k W^{a}_{\tilde{k}k}h^{t}_{jk} h^{t}_{lk}\right] A_{aijl} .
\end{equation}
Note that the nearest neighbor list used in the angular messages, $\tilde{N}(i)$, can differ from that used in the radial messages, $N(i)$, as the cutoff used for each is allowed to differ. In practice we use cutoffs of 5.2 \AA\ and 3.5 \AA\ for the radial and angular messages, respectively. As it was previously seen to improve performance in QRNN, we additionally include a charge-weighted radial AEV vector (QRAEV), which effectively encodes the coulombic interaction between charged atomic centers,
\begin{equation}
    q\tilde{R}_{ia} = \sum_j^{N(i)}q_iq_jR_{aij} .
\end{equation}
The QRAEV takes the same form as the radial messages $\tilde{R}_{i\tilde{k}a}$, using the partial charges as another node feature, and from now on we include it implicitly in $\tilde{R}_{i\tilde{k}a}$. The aggregated messages are then combined with the node features to predict an updated NFV and effective electronegativity via a trainable neural network
\begin{equation}
(\textbf{h}^{\prime \ t+1}_i, \chi_{eff,i}) = \mathcal{F}_{int}^{t}\big(q_i^t, h_{ik}^t, \tilde{R}_{i\tilde{k}a}^t,\tilde{A}_{i\tilde{k}a}^t\big),
\end{equation}
where $\mathcal{F}$ denotes an MLP with no activation function applied on the last layer. Note that the atomic charges, $q_i$, are initialized to zero on the first interaction block. 

After each message-passing iteration, MPNICE predicts equilibrated partial charges for each atom via an approximation to the Qeq method \cite{jacobson2022transferable}. The effective electronegativity of each atom $\chi_{eff,i}$ is predicted alongside the node features, $\textbf{h}^{\prime \ t+1}_i$, as an additional element in the output vector. The charge on atom $i$ is then updated by solving the analytic equation
\begin{equation}
q_i = -\frac{\chi_{eff,i} - \lambda}{J_{ii}},
\end{equation}
\begin{equation}
\lambda = \frac{Q_{tot} - \sum_i\frac{\chi_{eff,i}}{J_{ii}}}{\sum_i\frac{1}{J_{ii}}},
\end{equation}
where $Q_{tot}$ is the total charge of the system and $J_{ii}$ is a self-interaction term determined by the atomic radii.  It should be noted that this procedure guarantees that the sum of atomic energies is equal to the net charge and also that charge is globally equilibrated. Finally, the NFV for the central atom is updated, using a learned linear mixing of the predicted NFV with the previous NFV,

\begin{equation}
    \textbf{h}^{t+1}_{i} = \textbf{h}^t_i + ( W \times \textbf{h}^{\prime \ t+1}_i) + B^t .
\end{equation}

For the final energy block, after T interaction blocks, radial and angular messages are formed in the same way as in the interaction blocks, but instead of predicting new node features and effective electronegativities, the final MLP instead predicts $E_i$,
\begin{equation}
    E_i = \mathcal{F}_{energy}\big(q_i^T, h_{ik}^T, \tilde{R}_{i\tilde{k}a}^T,\tilde{A}_{i\tilde{k}a}^T\big).
\end{equation}
The final energy is then computed as a sum of atomic energy contributions, in addition to explicit contributions from electrostatics calculated from the last predicted partial charges, as well as, optionally, a Grimme D3 dispersion term.~\cite{Grimme20103} In the case that multiple levels of theory are being trained to, all layers but the final layer of $\mathcal{F}_{energy}$ are shared, with the intermediate representation $\textbf{h}_i^f$ being used as input to separate final layers for each output head.
\begin{equation}
    \textbf{h}_i^f = \mathcal{F}_{energy}\big(q_i^T, h_{ik}^T, \tilde{R}_{i\tilde{k}a}^T,\tilde{A}_{i\tilde{k}a}^T\big),
\end{equation}
\begin{equation}
    E_i(PBE) = \mathcal{F}_{PBE}\big(\textbf{h}^f_{i}),
\end{equation}
\begin{equation}
    E_i(\omega B97X) = \mathcal{F}_{\omega B97X}\big(\textbf{h}^f_{i}).
\end{equation}

\section{Results}\label{sec:results}

\subsection{Organic models}\label{sec:organic}
We train and test three models specific to tasks for organic molecules. \ORG\ is a multitask model trained to the SPICE dataset ($\omega$B97M-D3BJ/def2-TZVPPD),\cite{eastman2023spice,eastman2024nutmeg} OrbNet Denali dataset ($\omega$B97X-D3/def2-TZVP),\cite{christensen2021orbnet} and a proprietary dataset of 11M finite structures at the $\omega$B97X-D3BJ/def2-TZVPD level. \ORGTB\ is a delta-learned\cite{jacobson2022transferable} model trained to correct GFN2-xTB\cite{bannwarth2019gfn2} energies for the same dataset. \CORG\ is a multitask model trained to the same dataset, alongside a set of 400K molecular crystals evaluated using PBE-D3. For this model we report errors using two output heads, where it is of interest. 
 The output head being currently used is indicated in parenthesis, for example, \CORG\ (PBE-D3) indicates the use of the PBE-D3 output head of this multi-task model.  Unless otherwise specified the $\omega$B97X-D3BJ/def2-TZVPD output head is used.  More detailed information on how these datasets were constructed as well as tables summarizing the datasets and models trained can be seen in sections \ref{sec:datasets} and \ref{sec:training}.

We will compare results to a few other recent transferable organic MLFF models where possible, either using reported results from the literature or open source implementations, where feasible.  The first is ANI-2x\cite{Devereux2020ani-2x}, the latest in the line of ANI models, which supports only neutral molecules and 7 elements.  Next, is a version of our prior method, QRNN, \cite{jacobson2022transferable} which is a multi-task model\cite{jacobson2023leveraging} trained to our prior dataset,\cite{stevenson2019sani, jacobson2022transferable} enhanced with clusters extracted from crystal structures\cite{zhou2025CSP} and the OrbNet Denali set mentioned above.  In addition, we compare to a delta-learned model with the same design and dataset, which we will refer to as QRNN-TB.  Both of these models utilize the $\omega$B97X-D3/def2-TZVP level of theory.  We additionally compare to a prioprietary model based on the equivariant MACE architecture called MACE-OFF23\cite{kovács2025MACEOFF} that only supports neutral systems.  Finally, perhaps the most interesting comparison is to the recently released AIMNet2 model\cite{anstine2024aimnet2}, which also supports ionic systems, has a similar design and has good coverage of the relevant element space for organic molecules.  Here, we will utilize rotamer scans and relative tautomer energies to probe the accuracy of the models for describing small organic molecules.  A recent trend in the area of transferable organic force fields is to apply such models to compute condensed phase properties.\cite{anstine2024aimnet2, kovács2025MACEOFF}  Along those lines we will inspect the accuracy of the models to rank order organic crystals, provide stable MD trajectories and accurate liquid densities as well as the condensed phase properties of liquid water.

\subsubsection{Rotamers}\label{sec:rotamers}

Rotamer scans, also known as torsion scans, are a useful test to evaluate the ability of an organic force field to predict conformational energies of small molecules.  Further, traditional lifescience force fields, such as OPLS,\cite{damm2024opls5} are often fine-tuned to reproduce the rotamer scan of DFT prior to being used in biological simulations.  One potential application of an organic MLFF is as a drop in replacement for DFT to provide training data for such fine-tuning; another is ranking of conformations.

\begin{table}[!htb]
    \centering
    \begin{tabular}{|p{5cm}|p{3cm}|p{3cm}|p{3cm}|}
    \hline
Model	&	Genentech RMSD (CC)	&	TorsionNet500 RMSD (LOT)	&	TorsionNet500 RMSD (CC)	\\
    \hline
QRNN	&	0.43	&	0.53	&	0.59	\\
    \hline
QRNN-TB	&	0.33	&	0.35	&	0.47	\\
    \hline
\ORG	&	0.24	&	0.33	&	0.43	\\
    \hline
\ORGTB	&	0.21	&	0.19	&	0.33	\\
    \hline
\CORG\ ($\omega$B97X head)	&	0.30	&	0.37	&	0.45	\\
    \hline
\CORG\ (PBE head)	&	0.49	&	0.59	&	0.67	\\
    \hline
AIMNet2	&	0.47	&	0.47\cite{anstine2024aimnet2}	&	0.62	\\
    \hline
ANI-2x	&	0.72	&	1.90\cite{anstine2024aimnet2}	&	1.34	\\
    \hline
Orbnet Denali	&		&	0.18\cite{anstine2024aimnet2}	&		\\
    \hline
MACE-OFF23(L)$^\dagger$	&		&	0.25\cite{kovács2025MACEOFF}	&		\\
    \hline
    \end{tabular}
    \caption{Overall RMSD in relative energies (kcal/mol) for two sets of organic rotamers.  The Genentech set\cite{Sellers2017Genentech} gives CCSD(T)/CBS estimates for 62 molecules whereas the TorsionNet500\cite{Brajesh2022TorsionNet} set is a diverse set of 500 molecules.  We report TorsionNet500 results against the level of theory (LOT) models were trained to\cite{kovács2025MACEOFF, anstine2024aimnet2} as well as DLPNO-CCSD(T)/CBS estimates.  $^\dagger$MACE-OFF23(L) reported error is mean barrier height errors, not mean RMSD}
    \label{tab:rotamers}
\end{table}

In order to enable comparison with other, existing transferable organic MLFFs we report errors relative to two datasets from the literature:  TorsionNet500\cite{Brajesh2022TorsionNet} and the Genentech torsion set.\cite{Sellers2017Genentech}  TorsionNet500 is a diverse set of fragments of 500 neutral organic molecules covering the elements H C N O S F Cl.  Originally, the reference data was computed at the B3LYP/6-31G(d, p) level, and other workers utilizing this test have recomputed the energies at the level of theory for which the models were trained.\cite{anstine2024aimnet2, kovács2025MACEOFF}  To be as consistent as possible with previous reports we here report the overall RMSD in relative energies versus the trained level of theory, for each respective model;  the zero of energy for each torsion scan is given by the geometry with the lowest energy at the reference level of theory.  The central column of Table~\ref{tab:rotamers} compares the RMSD over the entire dataset for our MPNICE models, previously reported QRNN models, and several models from the literature.  We see that the delta-learned models give the highest accuracy, with Orbnet Denali (a delta-learned model utilizing QM features)\cite{christensen2021orbnet} giving the best results (0.18 kcal/mol) closely followed by \ORGTB\ (0.19 kcal/mol).  Following this, the direct learned models are lead by MACE-OFF23(L) (0.25 kcal/mol), \ORG\ (0.33 kcal/mol), \CORG\ ($\omega$B97X-D3BJ head) (0.37 kcal/mol), and AIMNet2 (0.47 kcal/mol), all able to reproduce torsion profiles with less than a half of a  kcal/mol accuracy.  We note however, that the MACE-OFF23 result is a different metric, namely the mean error in barrier height, and cannot be strictly compared. Unfortunately we are not able to recompute the errors here and only report what is publicly available as this model is proprietary.

We have also computed DLPNO-CCSD(T)/CBS estimates for the torsion curves in the TorsionNet500 dataset (details in section S7) as a high level benchmark.  The overall RMSD versus these values for our models and public models are given in the third column of Table~\ref{tab:rotamers}.  The accuracy here mirrors the accuracy versus the trained level of theory, but with deviations typically $\simeq$0.1 kcal/mol higher.  Notably, the \ORGTB\ model is able to reproduce these high level rotamer energies to 0.33 kcal/mol and our direct learned model can reproduce these energies to about 0.43 kcal/mol.  Interestingly ANI-2x appears to benefit from some cancellation of error, giving a lower error relative to DLPNO-CCSD(T)/CBS whereas AIMNet2 falls inline with the other models, delivering a 0.15 kcal/mol higher error at 0.62 kcal/mol.  We have also provided errors relative to CCSD(T)/CBS results for the Genentech rotamer set in column 1, which follows similar trends to the more diverse TorsionNet500 set.  We attribute the higher errors for TorsionNet500 to this increased diversity, not the slightly different level of theory.

\begin{figure}
    \centering 
    \caption{Violin plots showing the distribution of relative energy RMSDs for separate torsion scans in the Torsion2000 test set are given in panel (d), the line shows the median RMSD error. Panels (a)-(c) highlight three examples from models that give overall good performance on torsion scans.  (a) shows a high error example for AIMNet2, (b) shows a high error example for \ORG\ and \ORGTB\ and (c) shows a randomly selected torsion scan with representative performance.}
    \includegraphics[width=\linewidth]{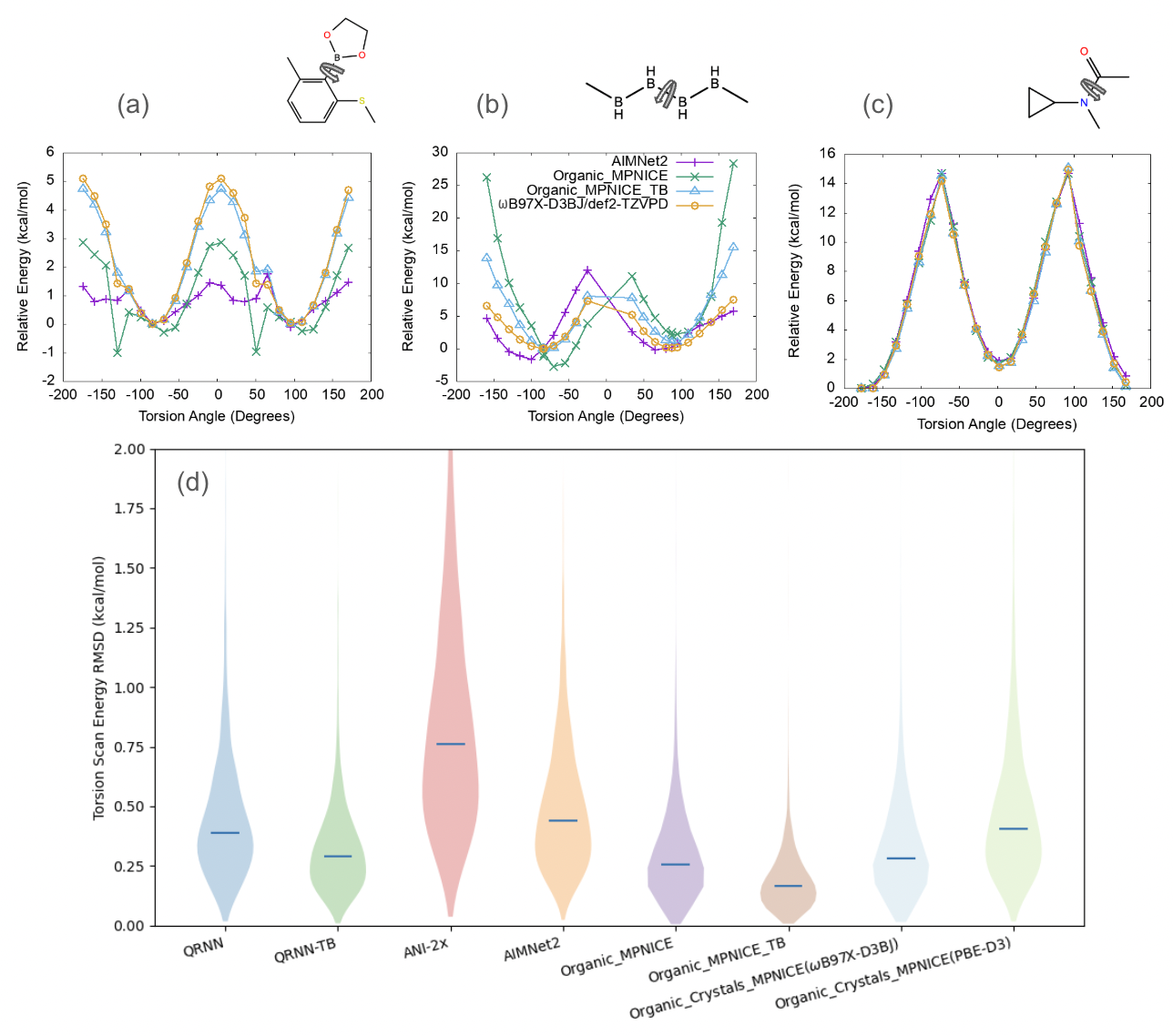} 
    \label{fig:torsiontest2000}
\end{figure}

The torsion tests described above give a good picture of the accuracy achievable with these models, however, they do not cover ionic molecules, or all of the most common elements found in drug-like molecules.  We have constructed another torsion test that covers such systems, which we refer to as Torsion2000, that contains about 2000 torsion scans labeled at the $\omega$B97X-D3BJ/def2-TZVPD level, see the supporting information for details.  Figure.~\ref{fig:torsiontest2000} (d) is a violin plot displaying the distribution of RMSD values.  For models only supporting a subset of the elements or only neutral molecules, the results are only accumulated for the supported elements and charge states. Amongst the models we are able to run, the \ORGTB\ model obtains the lowest median RMSD error (0.17 kcal/mol) followed by the \ORG\ model (0.26 kcal/mol).  AIMNet2 yields a higher median error than the MPNICE models (0.44 kcal/mol), but again strongly outperforms ANI-2x (0.76 kcal/mol).  However, it should be noted that AIMNet2 and ANI-2x are trained to different levels of theory and would be expected to have slightly higher error relative to this reference.

As can be seen in the violin plots all of the models have fairly long high error `tails' indicating a small number of high error cases.  Defining a high error case as one with an error of greater than 1 kcal/mol (chemical accuracy) we find that \ORGTB\ has the least such cases, with only 8, followed by \ORG\ with 24.  Excluding models that do not support all elements in this test the next highest number of high error cases in AIMNet2 with 118, followed by \CORG\ models.  Two such high error cases are shown in panels (a) and (b) in Figure~\ref{fig:torsiontest2000}, whereas panel (c) shows a randomly selected case with an error which is more representative of the set.  Panel (b) shows the highest error \ORGTB\ case, which also happens to have the highest error of \ORG\ and AIMNet2.  This is somewhat unsurprising as it is a molecule we would expect not to be representative of molecules selected from a database of druglike molecules, which these models are generally trained to.  AIMNet2 gives an error of 2.27 kcal/mol, much lower than the errors for both \ORGTB\ (3.19 kcal/mol) and \ORG\ (7.8 kcal/mol).  \ORGTB\ appears to give a reasonable reproduction of the minima in this case but makes very large errors on the barrier height for rotation.  Panel (a) shows the second highest error case for AIMNet2, another Boron containing molecule, for which AIMNet2 appears to perform qualitatively poorly, with a very flat PES.  In this case \ORGTB\ performs very well and \ORG\ appears to be almost an average of the two other models.  

\subsubsection{Tautomers}\label{sec:tautomer}

\begin{table}[!htb]
    \centering
    \begin{tabular}{|p{5cm}|p{3cm}|p{3cm}|}
    \hline
Tautobase (1499)	&	MAE (kcal/mol)	&	R$^2$	\\
    \hline
ANI-2x	&	2.90	&	0.78	\\
    \hline
QRNN	&	1.21	&	0.95	\\
    \hline
QRNN-TB	&	0.93	&	0.97	\\
    \hline
AIMNet2	&	0.86	&	0.98	\\
    \hline
\ORG	&	0.49	&	0.99	\\
    \hline
\ORGTB	&	0.29	&	1.00	\\
    \hline
\CORG\ ($\omega$B97X-D3BJ)	&	0.54	&	0.99	\\
    \hline
\CORG\ (PBE-D3)	&	1.84	&	0.89	\\
    \hline
    \end{tabular}
    \caption{The RMSD and $R^2$ values over a set of 1499 relative tautomer energies from the Tautobase dataset.  Reference energies are computed at the $\omega$B97X-D3BJ/def2-TZVPD level.}
    \label{tab:tautobase}
\end{table}
In addition to conformational energies, the prediction of relative tautomer energies is an interesting and useful task.  The need to rank order many tautomeric forms appears in computational workflows that estimate the pKa of organic molecules.  There is also much interest in predicting the most stable tautomeric forms of molecules in solution or in a protein environment with important applications in drug discovery research.\cite{dhaked2024impact}  

\begin{figure}[!hbt]
    \centering 
    \includegraphics[width=0.6\linewidth]{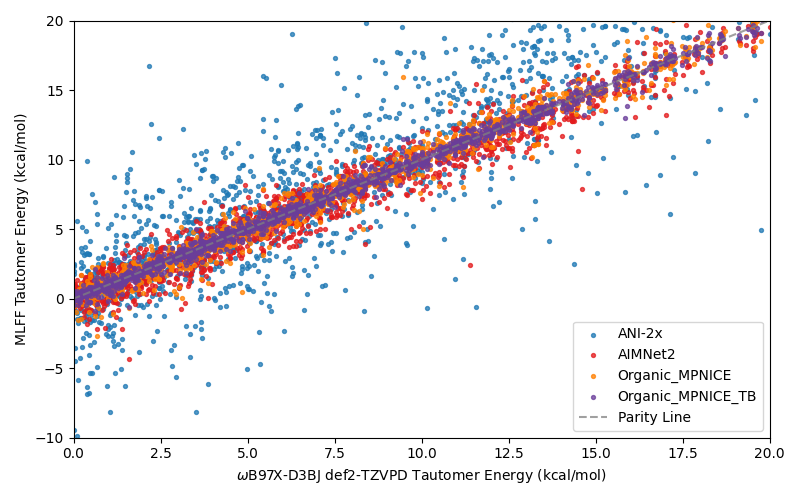} 
    \caption{Correlation plot showing the lower energy range of relative tautomer energies from the Tautobase set.}
    \label{fig:tautobase}
\end{figure}

Table~\ref{tab:tautobase} reports the overall RMSD and $R^2$ for the relative tautomer energies of a set of 1499 tautomer pairs (see the supporting information for details).  Here we see that the MPNICE models are strongly improved over prior QRNN models.  We find that \ORGTB\  and \ORG\ are able to reproduce these relative tautomer energies to within 0.5 kcal/mol whereas AIMNet2 is able to outperform our prior delta learning model and predict relative tautomer energies within chemical accuracy (1.0 kcal/mol).  We find ANI-2x performs relatively poor at this task (2.9 kcal/mol).  Owing to the large dynamic range of these relative tautomer energies (see Fig.~\ref{fig:tautobase}) the correlation of all models with reference results are fairly strong, varying from 0.78 for ANI-2x to an almost perfect correlation for \ORGTB.

\subsubsection{Organic Crystal Structures}\label{sec:CSP}

Many organic molecules are able to crystallize in different forms, called crystal polymorphs.\cite{Bernstein2020}  Organic crystal structure prediction (CSP) is an important computational workflow which seeks to elucidate the low energy crystal forms that can be observed experimentally and reduce the risk of so-called late appearing polymorphs.\cite{Bucar2015}  Such computations generally occur in two, potentially coupled stages;  the first stage produces a large number of proposed crystal structure candidates (packing search) while a second stage ranks the proposed structures according to energy (ranking).  While periodic PBE-D3 ranking has evolved into the workhorse of ranking, MLFF is emerging as a popular and promising method to reduce the cost of the ranking procedure.\cite{Egorova2020, McDonagh2019, Musil2018, Taylor2025, Wengert2021, Kadan2023, Butler2024}  Several participants in the recent 7th blind test utilized MLFF,\cite{Hunnisett2024BlindTestSearch,Hunnisett2024BlindTestRanking} in addition, we  have also incorporated our prior QRNN multi-task model into an end-to-end workflow, using both the pretrained model and a fine-tuning procedure to improve the robustness of ranking.\cite{zhou2025CSP}  Here, we will perform initial tests to evaluate the ability of MPNICE to rank order organic crystal structures and work as a drop in replacement of PBE-D3. Unfortunately, we are not aware of an implementation of GFN2-xTB that is suitable to compute energetics of such periodic systems and thus we do not perform \ORGTB\ predictions for these tests.

\begin{table}[H]
    \centering
    \begin{tabular}{|p{5cm}|p{2cm}|p{2cm}|}
    \hline
Model	&	Energy MAE (kcal/mol)	&	Density MAE (g/cm$^3$)	\\
    \hline
QRNN	&	187.8	&	0.052	\\
    \hline
\ORG	&	1.4	 &	0.070	\\
    \hline
\CORG\ ($\omega$B97X)	&	1.7	&	0.046	\\
    \hline
\CORG\ (PBE)	&	3.7	&	0.064	\\
    \hline
ANI-2X	&	4.9	&		\\
    \hline
MACE-OFF23 (S)$^\dagger$	&	3.1	&		\\
    \hline
MACE-OFF23 (M)$^\dagger$	&	1.8	&		\\
    \hline
MACE-OFF23 (L)$^\dagger$	&	1.7	&		\\
    \hline
    \end{tabular}
    \caption{Errors for lattice energies on the X23b test.   $\dagger$\ Errors for sublimation enthalpies on the same set, as reported by Kov\'acs \emph{et al.}\cite{kovács2025MACEOFF}  }
    \label{tab:x23b}
\end{table}

As an initial test of the ability of models to account for the interactions in organic crystals we evaluate lattice (cohesive) energies and mass densities in the X23b test set.  This is a set of 23 crystals of small, relatively rigid organic molecules for which experimental volumes and sublimation enthalpies are available.  The X23 dataset was originally constructed by Reilly and Tkatchenko\cite{Reilly2013X23} and later Dolgonos, Hoja and Boese\cite{Dolgonos2019X23b} revised reference values to remove thermal and zero-point effects, in an effort to make the references easier to compare to.

For all of the models reported here we have performed lattice optimizations and Table~\ref{tab:x23b} summarizes the errors in lattice energy and density, whereas figure S1 displays the correlation in lattice energies, as predicted by the models.  We have also listed the errors reported for MACE-OFF23 models and ANI-2x, as reported by Kov\'acs \emph{et al.}\cite{kovács2025MACEOFF}  Here, we report on the lattice energy of the solid, the cohesive energy normalized by the number of molecules, whereas Kov\'acs \emph{et al.} report the sublimation enthalpy, which explicitly takes into account the thermal effects.  We would expect the error on the sublimation enthalpy to be higher due to added errors in the phonon frequencies in the crystal.  We find that QRNN has an incredibly high error due to it's inability to reproduce the linear geometry of the CO$_2$ molecule.  Excluding this system the MAE in lattice energy for QRNN is 2.5 kcal/mol, which is quite a bit higher than the MPNICE models.  We find that both of the \ORG\ and \CORG\ ($\omega$B97X-D3BJ) models give good results for lattice energies and appear to outperform the largest MACE-OFF23 models.  Interestingly, adding periodic PBE-D3 data decreases the accuracy of the lattice energies and the PBE-D3 output head of \CORG\ gives quite poor lattice energies.  However, this output head has not been trained to any finite systems and may indicate poor reproduction of gas phase energies.

\begin{figure}
    \centering 
    \includegraphics[width=1.0\linewidth]{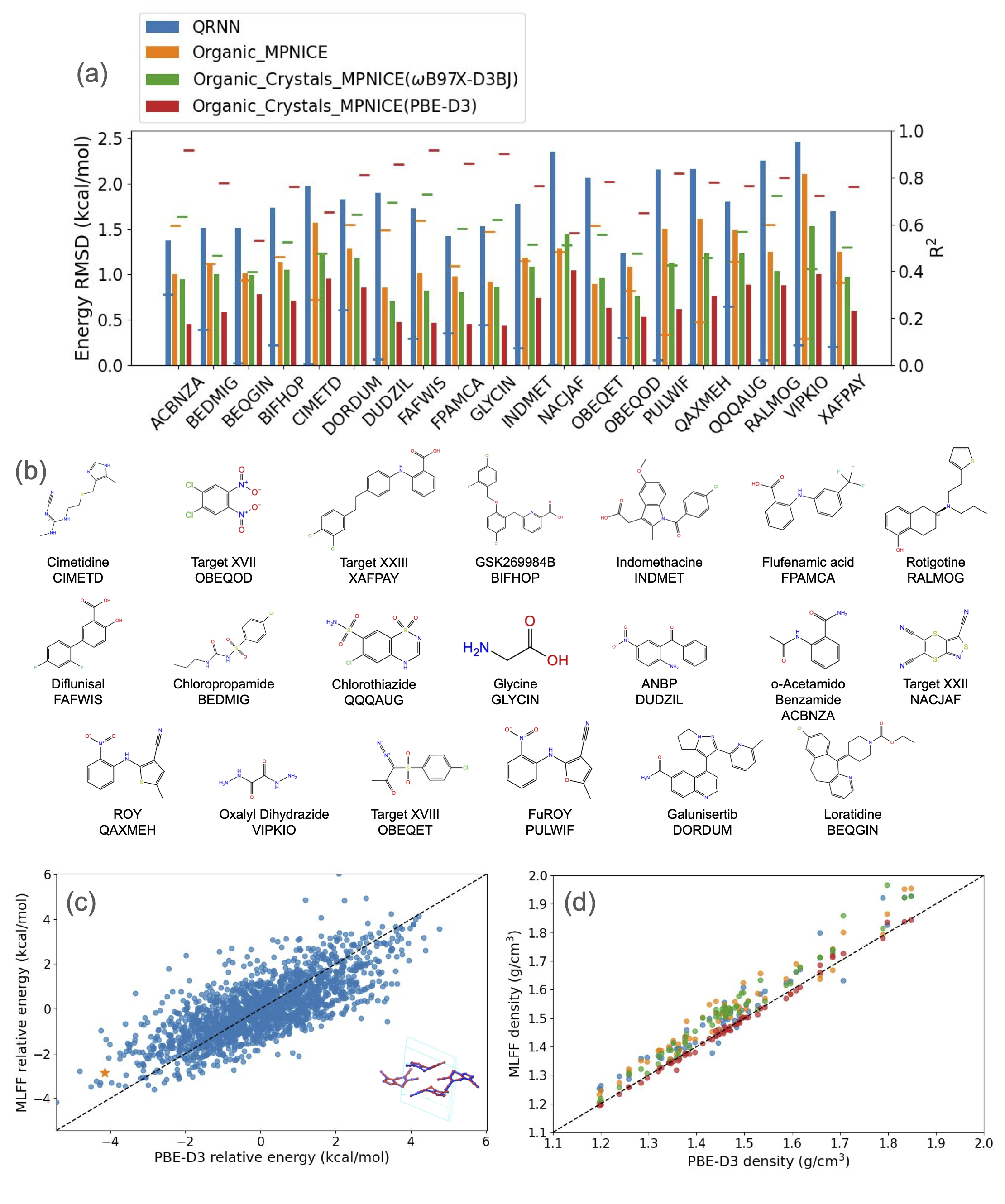} 
    \caption{\scriptsize Highlighted results for organic crystal structure ranking.  Panel (a) shows the RMSD in relative energy between MLFF models and loosely optimized PBE-D3 structures in kcal/mol (left axis) as bars and also a horizontal line indicating the $R^2$ value (right axis).  Panel (b) gives the common name and CCDC ref code for the systems considered.  Panel (c) shows the correlation of PBE-D3 relative energies with \CORG(PBE-D3) for the worst performing case of panel (a) (NACJAF).  The known experimental form is given as an orange star and an overlay of the optimized PBE-D3 and MLFF geometries is given in an inset.  Panel (d) shows the correlation of mass densities between MLFF models and tightly optimized PBE-D3 structures of 64 Z'=1 experimental forms corresponding to the systems studied here.  Common names and chemical structures for these systems are given in the supporting information. }
    \label{fig:CSP_results}
\end{figure}

Next, we will more directly interrogate the ability of the QRNN and MPNICE models to rank polymorphs.  We perform two tests:  First, we inspect the correlation of single point energies with loosely optimized PBE-D3 crystals; second, we inspect the consistency of densities with tightly optimized PBE-D3 crystals.  The CCDC (Cambridge Crystallographic Data Centre) ref codes of these systems are given in Figure~\ref{fig:CSP_results}, panel (b).  The CCDC ref codes for all 65 systems we previously studied were explicitly avoided in the generation of training data and so these systems serve as true tests for prospective use. Panel (a) of Figure.~\ref{fig:CSP_results} displays the RMSD of relative polymorph energies. When comparing relative energies, the RMSD can be particularly sensitive to the reference energy used, as such, we use the mean energy per molecule as the zero, and additionally report the coefficient of determination.

Perhaps unsurprisingly, the RMSD errors are lowest with the PBE-D3 output head of the \ORG\ model, with a mean RMSD of only 0.7 kcal/mol, followed by the $\omega$B97X-D3BJ head of the same model (1.0 kcal/mol), the \ORG\ model (1.2 kcal/mol) and finally our prior QRNN model (1.8 kcal/mol).  Importantly, errors on the systems with very high RMSD with QRNN are eliminated with the MPNICE models, particularly the PBE-D3 output head.  With the PBE-D3 output head, all systems have RMSD values less than about 1.0 kcal/mol and an $R^2$ value of at least 0.53.  Panel (c) of Figure.~\ref{fig:CSP_results} shows the correlation of what we judge to be the worst case, NACJAF, or Target XXII, a challenging case from a past blind test challenge.  Even though this is the worst case with relatively low correlation, the known experimental system (highlighted as a star) is ranked quite low in energy and the relative polymorph energies still show a strong correlation with PBE-D3.  The inset shows the overlap of the \CORG(PBE-D3) optimized structure and the tightly optimized structure with PBE-D3 (see below).

While energy correlation with PBE-D3 indicates that the \CORG\ model can be used as an effective filter for further structure refinement it is also useful to know how closely the fully optimized structures are to the PBE-D3 structures.  Figure~\ref{fig:CSP_results}, panel (d) shows the correlation of the mass density of 64 experimentally observed polymorphs corresponding to the 20 systems studied here, relative to tightly optimized PBE-D3 geometries (max force component is less than 0.03 eV/\AA).  We again find that the \CORG\ model, in combination with the PBE-D3 head is highly consistent with PBE-D3, giving a mean relative error in density of less than 0.5\%.  The other models trained to range-separated functionals all give consistently higher densities by about 3.5\% on average.  Due to this good agreement, the PBE-D3 head of the \CORG\ model could serve not only as a filter for PBE-D3 refinement of structures, but pre-optimization with this model may significantly reduce the burden of optimization and give highly consistent structures.  We also find that this model can be efficiently fine-tuned to specific molecules, which will be described in detail in an upcoming report.

\subsubsection{Molecular Dynamics}\label{sec:organic_MD}

While several transferable organic MLFFs have been demonstrated to accurately reproduce the conformational energetics of small molecules in a zero-shot scenario, some workers have pointed to the difficulty of zero-shot, stable, molecular dynamics simulations.\cite{Fu2023}  We and others, have previously demonstrated that MLFF models trained to clusters of specific systems are able to accurately reproduce experimental densities and other import liquid properties.\cite{dajnowicz2022high}   Here, we test the ability of transferable organic MPNICE models to generate stable trajectories for single component liquids and also reproduce the densities of such fluids near room temperature.  Recently, the MACE-OFF23 models have also been shown to yield reasonable density predictions,\cite{kovács2025MACEOFF} and for at least one system, the AIMNet2 model was shown to yield a stable condensed phase MD trajectory.\cite{anstine2024aimnet2}

For this test we have chosen a diverse set of 62 molecules that we use to construct single-component disordered systems for liquid simulation.  The systems chosen cover common organic laboratory solvents, solvents commonly used in liquid Li-ion battery electrolyte formulations and a diverse subset of those systems studied by Kov\'acs \textit{et al}.\cite{kovács2025MACEOFF}  These systems cover non-polar molecules as well as a diverse set of common organic functional groups as well as halogenated systems covering fluorinated, chlorinated and brominated species.  See the supporting information for details.

\begin{figure}[!hbt]
    \centering 
    \includegraphics[width=0.8\linewidth]{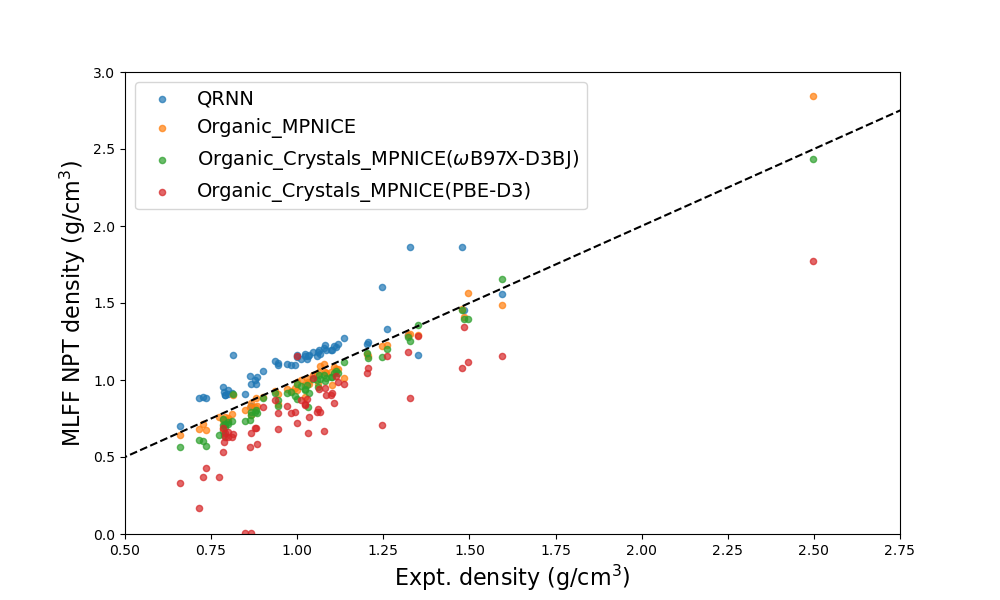} 
    \caption{Equilibrated densities for a set of 62 common organic solvents versus experiment, in. The full list of solvents and densities can be seen in table S2. Note that the QRNN model used does not support Br containing solvents.
    }
    \label{fig:liquid_solvent_densities}
\end{figure}

We find that with the MPNICE models, all simulations are stable and produce fairly reliable density estimates.  This is in contrast to our prior model, QRNN, where an ensemble of five models is required to achieve stable trajectories.  A correlation of the experimental and predicted densities is shown in Figure.~\ref{fig:liquid_solvent_densities}.  We see that the \ORG\ model provides the best average densities with an average error of about 4\% whereas the \CORG\ model with the $\omega$B97X-D3BJ head produces slightly underestimated densities and a 7\% average percent error.  The PBE-D3 output head of the \CORG\ performs surprisingly poorly, generally underestimating densities with some very large errors. In three cases, \CORG\ (PBE-D3) underestimates the boiling point and vaporizes the system (see SI). Additionally, the QRNN method gives systematically too high densities and an average percent error of 16\%.  Our best performing model, \ORG\ performs a slightly better than the average errors reported for the MACE-OFF23 model, for which Kov\'acs \emph{et al.}\cite{kovács2025MACEOFF} report an average error, on similar systems, of 0.09 g / cm$^3$ whereas here we see an error of 0.04 g / cm$^3$. Overall, we find it very encouraging that these new models, MACE-OFF23 and \ORG\ are able to perform zero-shot MD simulations and reproduce experimental densities over a fairly diverse range of molecular liquids.  

\subsubsection{Water Properties}\label{sec:water}

\begin{table}[H]
    \centering
    \begin{tabular}{|c|p{2cm}|p{3.3cm}|}
    \hline
	&	Experiment	&	\ORG	\\
    \hline
Density (g/cm$^3$)	&	0.998	&	0.933	\\
Diffusion coefficient (m$^2$/s)	&	2.30E-09	&	2.37E-09	\\
Hvap (kJ/mol)	&	40.65	&	48.52 \\
Cv (J/g/K)	&	4.18	&	4.70		\\
Hydration free energy (kcal/mol)	&	-6.31	&	-6.25		\\
    \hline
    \end{tabular}
    \caption{Water properties at 25\degree C calculated via \ORG\ versus experiment. Experimental water hydration free energy is obtained from Kelly, Cramer and Truhlar, \cite{kelly2005sm6} while other experimental values are obtained from the CRC handbook.\cite{lide2004crc}}
    \label{tab:water}
\end{table}

\begin{figure}[H]
    \centering 
    \includegraphics[width=\linewidth]{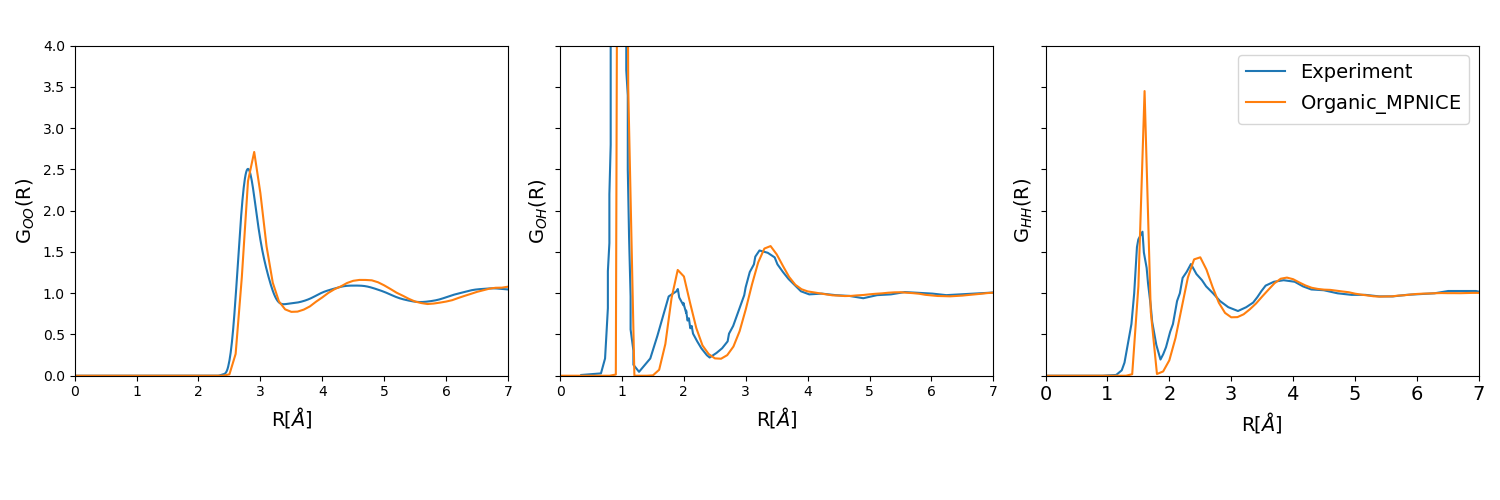} 
    \caption{Structure Factors for liquid water for \ORG\ versus experiment. Experimental values were taken from Soper \emph{et al.}~\cite{soper2000radial}. }
    \label{fig:water_RDF}
\end{figure}

We next turn our attention to the condensed phase properties of perhaps the most studied liquid, water.  We will restrict our attention to the one model (\ORG) which we have found performs best for liquid properties.  Table~\ref{tab:water} lists the average mass density, diffusion coefficient, heat of vaporization, constant volume heat capacity, and free energy of solvation (excess chemical potential) predicted by our model.  The density is underestimated by 0.06 g / cm$^3$, slightly higher than the average error of other liquids studied in section~\ref{sec:organic_MD}.  The diffusion coefficient and solvation free energy are surprisingly accurate whereas the heat of vaporization and heat capacity are reasonable approximations to the experimental values.  Nuclear quantum effects are known to impact the prediction of some of these properties, which are not accounted for in these classical estimates.  The OO, OH, and HH radial distribution functions ($g(r)$) are shown against  experimental neutron scattering results in Figure~\ref{fig:water_RDF}.  Overall, the agreement is good, however, the first oxygen peak in the OO distribution is at a slightly too large distance, reflecting the too low density estimate and the distributions are slightly over-structured.  Interstingly, Kov\'acs \emph{et al.} report that the MACE-OFF series of models report too high densities of water (12-20\%), depending on the particular hyperparameters, and that models with increased receptive fields were required to obtain very close density estimates.  The authors attribute this to the need to account for longer-ranged interactions.  Our models incorporate long-range interactions through fractional charge based coulomb interactions and long-range dispersion interactions, which, if we accept their hypothesis, allows us to give reasonable agreement with the experimental density.  Generally, the structure, energetic properties, and dynamics of liquid water appear to be well accounted for in this model.

\subsubsection{Application Highlight: Vertical Ionization of Small Molecules}\label{sec:IPEA}
Electron transfer and redox reactions hold fundamental significance in understanding chemical reactivity in drug discovery and materials science. Arguably the simplest of these cases to test is single vertical ionization potential (IP) and electron affinity (EA), which is the change in energy upon adding or removing an electron from the neutral state without allowing the geometry to relax,
\begin{equation}
    \text{IP} = E_\text{cation} - E_\text{neutral},
\end{equation}
\begin{equation}
    \text{EA} = E_\text{neutral} - E_\text{anion}.
\end{equation}
The molecular geometry does not change, and as such, vertical ionization energies test the model's ability to predict the properties of a species based on the charge alone, which is intractable without some form of charge feature.\cite{jacobson2022transferable} 

\begin{table}
    \centering
    \begin{tabular}{|c|c|c|c|c|c|} \hline 
         Model&  IP&  EA&  Cation&  Neutral& Anion\\ \hline 
         MPNICE from scratch&  3.87&  4.09&  4.37&  3.96& 4.8\\ \hline 
         MPNICE fine-tuned from \ORG\ &  3.01&  2.79&  2.97&  2.08& 3\\ \hline 
         AIMNet-NSE (ens5)&  2.7&  2.4&  4&  3.4& 3.8\\ \hline
    \end{tabular}
    \caption{Validation RMSEs in kcal/mol for ionization potential (IP), electron affinity (EA), and total energies for each subset of charged species (cation, neutral, anion) in the CHEMBL20~\cite{zubatyuk2021teaching} test set.}
    \label{tab:IP_EA_performances}
\end{table}

In Table~\ref{tab:IP_EA_performances}, we compare against AIMNet-NSE\cite{zubatyuk2021teaching}, an architecture that has seen success in modeling these reactions for drug-like molecules. As the AIMNet-NSE training set is not publicly available, the comparison against the results reported by the AIMNet-NSE references is not between equivalently trained models, but rather to show that MPNICE is able to train to a similar dataset with comparable results. AIMNet-NSE was trained to the IONS-12 dataset, which has 6.44M points and was tested on IONS-16 (300k points) and CHEMBL-20 (800 points), which are datasets of increasing maximum number of heavy atoms (12, 16, and 20 respectively). For MPNICE, we start with the \ORG\ model and finetune to IONS-16, testing on CHEMBL-20. The IONS-16 and CHEMBL-20 sets are publicly available\cite{zubatyuk2021teaching_database} and used as-is. Compared to the from scratch (trained to IONS-16 only) model, there are noticeable improvements in performance on the order of more than 1 kcal/mol in energy. The single model finetuned MPNICE test set performance is comparable to that of the AIMNet-NSE ensemble model.

\subsection{Inorganic models}\label{sec:inorganic}

We train inorganic MPNICE models to reproduce the PBE settings used in \MPTRJ, for which many model architectures have been previously trained and benchmarked.\cite{deng2023chgnet,chen2022universal,batatia2023foundation,fu2025learning,barroso2024open,park2024scalable,choudhary2021atomistic,bochkarev2024graph, merchant2023scaling, neumann2024orb, zhang2024dpa} As \MPTRJ\ does not contain atomic partial charge data, we ran GFN1-xTB on the \MPTRJ\ set and train to reproduce GFN1-xTB atomic partial charges in addition to the \MPTRJ\ PBE energies, forces, and stress. We refer to this model as \MPT.

In addition to \MPTRJ, the OMAT24 dataset was recently released as an attempt to increase the sampling of datapoints far from equilibrium.\cite{barroso2024open} To test the impact of including such data, we additionally trained an MPNICE model on the \emph{ab initio} molecular dynamics (AIMD) subset of OMAT24 run at 3000K, containing 13.9M datapoints, which we refer to as OMAT24$_a$; the model trained to this is referred to as \OMA. As OMAT24 has slightly incompatible atomic energies versus \MPTRJ, primarily due to a difference in pseudopotentials used, we took \OMA\ and trained for an additional 50 epochs on the \MPTRJ/xTB set in order to directly benchmark versus materials project DFT values. This model is referred to as \MAT. Additional details on training for these three models can be found in section \ref{sec:training}.

\subsubsection{Energy Ranking}\label{sec:energyranking}
We first evaluate the accuracy of these models at equilibrium geometries, which primarily involves evaluating the relative stability of different phases with respect to structural and elemental composition. Arguably the most fundamental task within this realm is to predict the relative energetics of monoelemental crystal structures. This is necessary to accurately predict the most stable crystal form for a given element, and thus obtain accurate bulk properties and formation energies. To test this, we extracted all optimized crystals from the Materials Project which consist of a single element. As none of the MLFFs we test explicitly include magnetic information, we do not include magnetic materials. Additionally, we screen out un-optimized structures and those for which PBE total energies are not available. This results in a set of 186 optimized monoelemental crystals, with some bias towards elements with a large variety of sampled structures, i.e. C, Si, and O (Table S1). In addition to the MPNICE models, we include results for two previously trained MLFFs from the literature, MatterSim-v1.0(1M)\cite{yang2024mattersim} and SevenNet-0\cite{park2024scalable}. These models were chosen for comparison due to their availability, strong performance on prior benchmarks (see Section \ref{sec:mech_phonon}), and competitive inference speed. SevenNet-0 is an equivariant model with seven message passing iterations, and was trained on \MPTRJ. MatterSim is an invariant model, but was trained on a significantly larger proprietary dataset\cite{yang2024mattersim}. Summarized results for single point calculations can be seen in Table \ref{tab:singleElementCrystal}. In addition to the energy MAE over the whole set, we report an average correlation R$^2$ value, which is intended to measure the ability of an MLFF to distinguish between distinct crystal structures with the same element. The correlation between MLFF energy and PBE energy upon linear regression is calculated separately for each element, and averaged over all elements with greater than two datapoints.

\begin{table}[!htb]
    \centering
    \begin{tabular}{|c|c|c|}
    \hline
    Model	&	Energy MAE (eV/atom)	&	Mean Energy R$^2$	\\
    \hline
    \OMA	&	0.082	&	0.87	\\
    \hline
    \MPT	&	0.026 &	0.91	\\
    \hline
    \MAT	&	0.022	&	0.92	\\
    \hline
    MatterSim	&	0.056	&	0.81	\\
    \hline
    SevenNet	&	0.037	&	0.78	\\
    \hline
    \end{tabular}
    \caption{Energy mean absolute errors (MAEs) and average R$^2$ for the full set of 186 monoelemental crystal single points. The average R$^2$ was calculated via calculating the line of best fit for each element, and averaging the resulting correlation R$^2$ over all elements with greater than two structures contained in the test set. The specific models used for MatterSim and SevenNet are MatterSim-v1.0.0-1M and SevenNet-0, respectively.}
    \label{tab:singleElementCrystal}
\end{table}

\OMA\ performs the worst with respect to single point total energies, potentially due to the differences in pseudopotentials used in Materials Project and OMAT24. However, it outperforms both MatterSim and SevenNet in the mean energy R$^2$, indicating that the model is still capable of distinguishing between structures with very slight structural differences. The two MPNICE models trained to \MPTRJ\ energies perform the best on the single point energy test, with \MAT\ slightly outperforming \MPT, with a 22 meV/atom MAE. SevenNet performs slightly worse, at 37 meV/atom MAE, but with significantly lowered mean R$^2$. MatterSim comes in last with double the MAE of \MAT\ at 56 meV/atom, but slightly improves upon the mean R$^2$ of SevenNet, suggesting the higher MAE is, similar to \OMA, due to slight differences in settings used to generate training data. All models display a strong correlation with \MPTRJ.

We then perform the same test, while allowing the atomic positions and lattice vectors to relax using the MLFF PESs (Table \ref{tab:singleElementCrystalOpt}). To prevent collapsing to the lowest energy space group, the symmetry was constrained within ASE. Even though the initial structures were already optimized using PBE, all models have a relatively large change in volume/atom, resulting in a significant increase in Energy MAE as compared to the optimized PBE values. The trends between models remain consistent, with \MAT\ achieving the best MAE for both volume and energy, followed by \MPT\ and \OMA. All models obtain relatively high overall MAE for volume upon optimization. Some elements are especially problematic for these MLFFs in their monoelemental form, including halogens (F, Cl, Br), alkaline earth metals (Ca, Sr, Ba, Ra), and some alkali earth metals (K, Rb).

\begin{table}[!htb]
    \centering
    \begin{tabular}{|c|c|c|c|}
    \hline
    Model	&	Volume MAE (\AA$^3$/atom)	&	Energy MAE (eV/atom)	&	Mean energy R$^2$	\\
    \hline
    \OMA	&	3.19	&	0.113	&	0.71	\\
    \hline
    \MPT	&	2.46	&	0.085	&	0.71	\\
    \hline
    \MAT	&	2.39	&	0.076	&	0.74	\\
    \hline
    MatterSim	&	3.33	&	0.091	&	0.64	\\
    \hline
    SevenNet	&	3.29	&	0.086	&	0.65	\\
    \hline
    \end{tabular}
    \caption{Volume and energy mean absolute errors (MAEs) and average R$^2$ for the full set of 186 monoelemental crystal optimizations. }
    \label{tab:singleElementCrystalOpt}
\end{table}

Many of the crystal structures in the monoelemental test are well represented in \MPTRJ, making this test decidedly in-distribution and potentially subject to overfitting. As inorganic MLFFs are often used to model interactions of basic systems such as these (e.g. Lithium metal/electrolyte interface), the ability of inorganic MLFFs to distinguish the energetics of fine structural changes of basic crystals is important enough to be benchmarked, if not prioritized. However, benchmarks for explicitly out-of-distribution (OOD) tests must also be used to detect overfitting and transferability.

\vspace{10pt}
\begin{figure}[hbtp]
    \centering 
    \includegraphics[width=\linewidth]{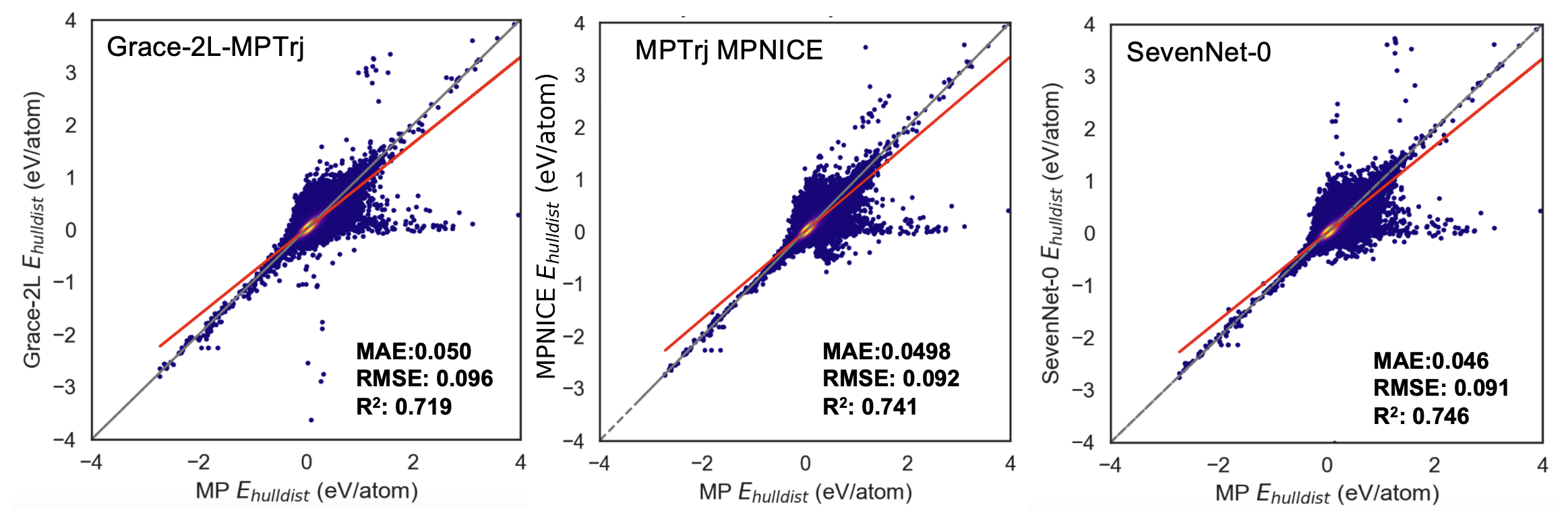} 
    \caption{DFT vs MLFF convex hull distance for \MAT, and the two closest models, Grace-2L and SevenNet-0}
    \label{fig:MatbenchParity}
\end{figure}

\vspace{10pt}
\begin{table}[hbtp]
\centering
\footnotesize
\begin{tabular}{|l|c|c|c|c|c|c|c|c|c|c|}
\hline
\textbf{Model} & Params & F1 & DAF & Prec & Acc & TPR & TNR & MAE & RMSE & R$^2$ \\
\hline
Grace-2L-MP$_{trj}$ & 15.3M & 0.685 & 3.779 & 0.630 & 0.885 & 0.749 & 0.912 & 0.050 & 0.096 & 0.719 \\
\hline
\MPT\ & 1.57M & 0.687 & 3.557 & 0.593 & 0.876 & 0.816 & 0.888 & 0.050 & 0.092 & 0.741 \\
\hline
\MAT\ & 2.7M & 0.713 & 3.704 & 0.617 & 0.887 & 0.843 & 0.896 & 0.044 & 0.088 & 0.764 \\
\hline
SevenNet-0 & 842.4 K & 0.720 & 3.884 & 0.647 & 0.894 & 0.811 & 0.912 & 0.046 & 0.091 & 0.746 \\
\hline
\end{tabular}
\caption{Matbench Discovery statistics for \MPT\ and models with similar performance}
\label{MatbenchStats}
\end{table}

 The Matbench Discovery framework provides a robust set of metrics to evaluate a model's ability to predict OOD thermodynamic stability for bulk inorganic crystals. The framework leverages DFT-calculated energies and convex hull distances from the WBM (Wang-Botti-Marques) dataset,\cite{wang2021predicting} which comprises approximately 257,000 structures. These structures are generated via chemical-similarity-based elemental substitutions of source structures from the Materials Project (MP). For an MLFF model to be effective in real-world materials discovery, it should correctly predict relative energies, thereby enabling the identification of ground-state thermodynamic stability for structures absent from the training data. The WBM test set is particularly well-suited to evaluate this capability, especially for models trained on the Materials Project (MPTrj) dataset.
When evaluated on Matbench Discovery, the \MPT\ model (Figure \ref{fig:MatbenchParity}(b)) demonstrates competitive performance among compliant models (i.e., models trained on the MPTrj dataset). All the metrics reported are on the full WBM test set, and not the subset of unique prototypes. \MPT\ obtains F1 score of 0.687, placing it just above GRACE-2L-MP$_{trj}$ (0.685) and  slightly behind SevenNet-0 (0.720). Interestingly, as seen in Table \ref{MatbenchStats}, \MPT\ surpasses its two closest competitors, SevenNet-0 and Grace-2L, in true positive rate (TPR)—its ability to correctly identify DFT-stable materials. However, its lower precision indicates a higher rate of false positives, suggesting that the hypothetical convex hull predicted by \MPT\ is systematically shifted downward compared to the MP convex hull. As a result, while the model achieves better classification of stable materials, it does so at the expense of slightly overpredicting stability for some structures.  The results for \MAT, being pretrained on the OMAT24$_a$ subset, are slightly improved versus \MPT\ on this benchmark, although the use of some OMAT24$_a$ data results in non-compliance with the matbench discovery benchmark.

\begin{figure}[!htb]
    \centering
    \includegraphics[width=0.6\linewidth]{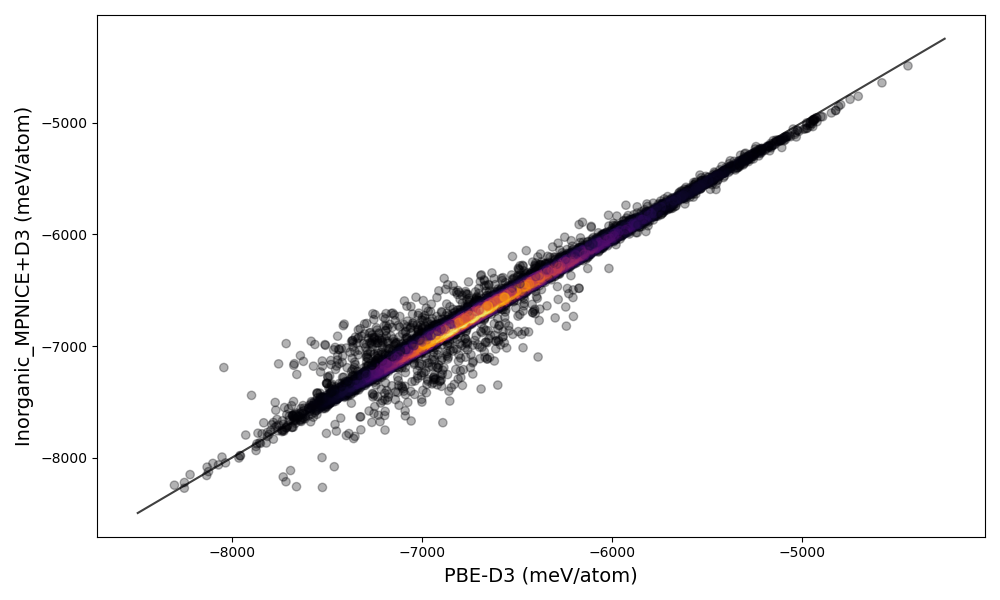}
    \caption{Parity plot between \MAT+D3 and PBE-D3 for the normalized total energies of $>$20k metal organic frameworks in the QMOF dataset.}
    \label{QMOF_parity}
\end{figure}

For an additional test of OOD energetic ranking, we ran \MAT\ on the QMOF dataset of over 20k optimized Metal Organic Framework structures,\cite{rosen2021machine,rosen2022high} for which the MACE$\_$MP$\_$0 \MPTRJ\ trained model reported a single point energy MAE of 33 meV/atom versus PBE-D3.\cite{batatia2023foundation} To reproduce PBE-D3 total energies, we add an additional Grimme D3 correction on top of the MPNICE PBE energies. \MAT\ performs similarly to MACE$\_$MP$\_$0, with a slightly lower MAE of 30 meV/atom. A parity plot of the \MAT+D3 energies and PBE-D3 for the QMOF dataset can be seen in Figure \ref{QMOF_parity}

\subsubsection{Mechanical Properties Near Equilibrium}\label{sec:mech_phonon}
\begin{figure}[!htb]
    \centering
    \includegraphics[width=\textwidth]{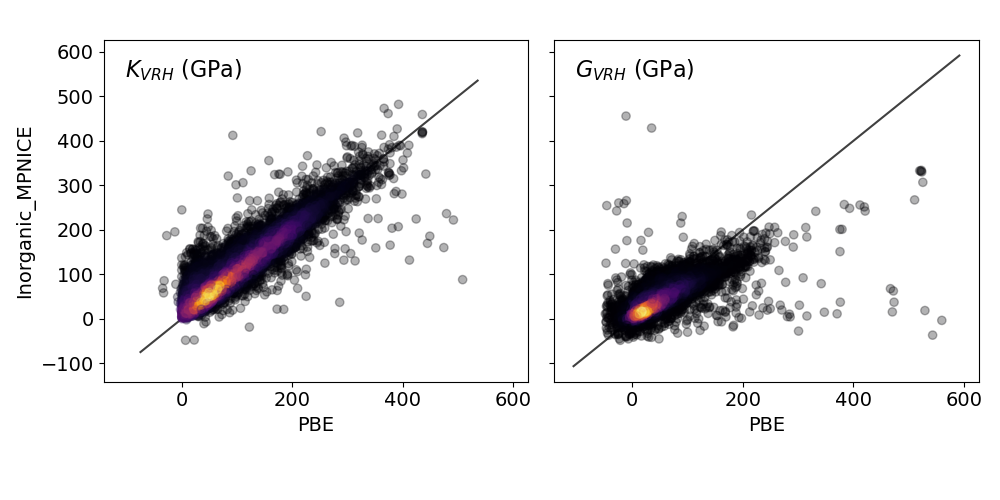}
    \caption{Parity plots for \MAT\ on the Materials Project elastic dataset showing the bulk modulus (left) and shear modulus (right) for 11.6k datapoints. Note that the plots exclude 2 and 82 outliers for which \MAT\ predicted a nonphysical bulk and shear modulus, respectively, of less than -50 or greater than 600 GPa.}
    \label{fig:mechanical_properties}
\end{figure}

In addition to ranking structures based on energetics at equilibrium, MLFFs are increasingly used to calculate mechanical properties of bulk crystals. These properties are typically calculated via finite difference around the equilibrium structure, and are a good measure of model accuracy immediately surrounding equilibrium. To this end we compiled a dataset from all datapoints in the Materials Project database which include bulk  ($K_{VRH}$) and shear moduli ($G_{VRH}$). Each structure was first optimized using the MPNICE model, before sampling 24 strained geometries, evaluating the resulting single point energies, and using these to fit the 21 parameter elastic tensor. This was then used to compute $K_{VRH}$ and $G_{VRH}$, following the procedure outlined in ref. \citenum{batatia2023foundation}. In all models there were some systems with very large errors in either $K_{VRH}$ or $G_{VRH}$, defined as either $<-50$ GPa or $>600$ GPa; these represented less than 1$\%$ of the test set for all models and were excluded from the statistics for each, seen in Table \ref{tab:elastic}.  Interestingly, \MAT\ performs the best for volume MAE, but exhibited the largest number of outliers in the shear modulus, 82. The inclusion of more data off of equilibrium via OMAT24$_a$ in \OMA\ and \MAT\ results in a significant reduction in the MAE for $G_{VRH}$ vs \MPT. Overall all models achieve reasonable performance, comparable to previous reports of MACE$\_$MP$\_$0 \cite{batatia2023foundation} on a similar benchmark. Parity plots of $K_{VRH}$, and $G_{VRH}$ for \MAT\ versus DFT can be seen in Fig. \ref{fig:mechanical_properties}. 

\begin{table}[!htb]
    \footnotesize
    \centering
    \begin{tabular}{|p{3cm}|p{1.45cm}|p{1.45cm}|p{1.45cm}|p{1.45cm}|p{1.45cm}|p{1.45cm}|p{1.45cm}|}
    \hline
Model	&	Volume MAE (\AA$^3$/atom)	&	$\#$ $K_{VRH}$ outliers	&	$K_{VRH}$ MAE (GPa)	&	$K_{VRH}$ R$^2$	&	$\#$ $G_{VRH}$ outliers	&	$G_{VRH}$ MAE (GPa)	&	$G_{VRH}$ R$^2$	\\
    \hline
\OMA	&	0.569	&	5	&	22	&	0.78	&	40	&	16	&	0.62	\\
    \hline
\MPT	&	0.555	&	0	&	28	&	0.86	&	74	&	37	&	0.45	\\
    \hline
\MAT	&	0.504	&	2	&	21	&	0.84	&	82	&	19	&	0.52	\\
    \hline
    \end{tabular}
    \caption{Statistics for the bulk modulus ($K_{VRH}$) and shear modulus ($G_{VRH}$) of a set of 11.6 thousand structures for MPNICE models. Systems tagged as outliers are excluded from their respective analyses.}
    \label{tab:elastic}
\end{table}

We next ran our MPNICE models on the recently reported PBE MDR benchmark on phonon and thermal properties,\cite{loew2024universal} for which results can be seen in Table \ref{tab:MDR_phonons}. 
It is clear that the inclusion of larger amounts of training data sampled away from equilibrium, as present in \OMA\ and the proprietary dataset of MatterSim, increases accuracy for these thermal properties. Interestingly, the benefit from this is significantly reduced upon subsequent training to \MPTRJ, and \MAT\ exhibits worsened MAE for thermal properties, even while obtaining a lower volume MAE. As noted in ref. \citenum{loew2024universal}, the ORB and OMat24 models, which directly predict gradients instead of calculating them via differentiation to save cost, are unable to reproduce thermal properties. \OMA\ and MatterSim both approach errors within the magnitude of discrepancy between PBE and PBEsol.

\begin{table}[!htb]
    \centering
    \begin{tabular}{|c|c|c|c|c|c|c|c|}
    \hline
Model	&	\% failure	&	E &	V &	$\omega$&	S	& F &	Cv \\
    \hline
ORB$^a$	&	0.82	&	31	&	0.082	&	291	&	421	&	175	&	57	\\
    \hline
OMat24$^a$	&	0.85	&	33	&	0.084	&	780	&	403	&	241	&	100	\\
    \hline
M3GNet$^a$	&	0.12	&	33	&	0.516	&	98	&	150	&	56	&	22	\\
    \hline
CHGNet$^a$	&	0.09	&	334	&	0.518	&	89	&	114	&	45	&	21	\\
    \hline
MACE$^a$	&	0.14	&	31	&	0.392	&	61	&	60	&	24	&	13	\\
    \hline
\textbf{\MAT	}&	0.09	&	39 &	0.348	&	55	&	49	&	19 &	12	\\
    \hline
SevenNet$^a$	&	0.15	&	31	&	0.283	&	40	&	48	&	19	&	9	\\
    \hline
\textbf{\OMA	}& 0.07		&	17 &	0.387	&	39	&	23	&	7 &	7	\\
    \hline
MatterSim$^a$	&	0.1	&	29	&	0.244	&	17	&	15	&	5	&	3	\\
    \hline
PBEsol$^a$	&		& 	&	1.283	&	33	&	25	&	10	&	5	\\
    \hline
    \end{tabular}
    \caption{MAE for the PBE MDR dataset, for Energy E (meV/atom),	Volume V (\AA$^3$/atom),	Maximum phonon frequency $\omega$ (K), Entropy	S (J/K/mol), Helmholtz Free Energy F (kJ/mol), and heat capacity Cv (J/K/mol). Results for models with $^a$ were taken from Ref. \citenum{loew2024universal} }
    \label{tab:MDR_phonons}
\end{table}

\newpage
\subsubsection{External Electric Fields}\label{sec:ext_fields}

Although the MDR set provides a good test for the higher order geometric response of inorganic MLFFs, the benchmark excludes non-analytic corrections (NAC) to the phonon band structure due to long-range electrostatics. Including such nonlocal coulomb effects is necessary to obtain qualitatively correct band structures for polar/ionic systems, and can affect relevant observables such as thermal conductivity.\cite{shafique2020effect} This correction is typically removed from large scale phonon benchmarks for MLFFs\cite{yu2024systematic} as universal MLFFs without internal charge representations cannot directly predict dielectric tensors.

Recently, multiple studies on machine learning the response to an external electric field have been reported, via training to a specific system using the external field as an equivariant feature.\cite{zhang2023universal, mao2024dielectric,falletta2024unified} As atomic partial charges are explicitly represented in the MPNICE architecture, we can in principle directly access linear response properties with respect to external electric fields.  We incorporate the external field into the neural Qeq as a linear perturbation of the effective electronegativity of each atom,

\begin{equation}
    \tilde\chi_{i} = \chi_{i} - \Sigma_{\alpha}E_{ext,\alpha}R_{i,\alpha},
\end{equation}
where $\chi_{i}$ is the effective electronegativity of the atom i predicted by the last interaction block, $\alpha$ refers to the cartesian coordinates \{x, y, z\}, $E_{ext,\alpha}$ is the external electric field along the $\alpha$ direction, and $R_{i,\alpha}$ is the cartesian coordinate of atom i. This technique has been used with some success with Qeq based classical force fields\cite{assowe2012reactive}. The perturbed effective electronegativities are then used to predict equilibrated charges, which are then used as input in the readout block of MPNICE and electrostatic energy. Note that the Ewald summation of the electrostatic energy is also perturbed by a linear external field. Response properties, including the dielectric susceptibility, Born effective charges ($Z^*_{i\alpha\beta}$), and piezoelectric tensor, are then accessible as second derivatives of the energy with respect to the external field, cartesian coordinates, lattice parameters, or combinations thereof, following the linear response theory of DFPT\cite{wu2005systematic}. These can be calculated for MLFFs either by finite difference or directly via automatic differentiation; we opt to use automatic differentiation for the results in this paper. While the fixed ion dielectric tensor ($\epsilon^\infty_{\alpha\beta}$) and $Z^*_{i\alpha\beta}$ are functionally second derivatives with respect to the energy, it is possible to refactor these into single automatic differentiation derivatives within our formalism, making use of the relation between the dipole moment and the energy,

\begin{equation}
    \mu_{\alpha} = -\frac{dE}{d\vec{E}_{el,\alpha}},
\end{equation}
where $\mu_{\alpha}$ is the dipole moment  and $\vec{E}_{el,\alpha}$ is the external electric field in cartesian direction $\alpha$. For a set of point charges $q_i$ at positions $R_{i\alpha}$, the dipole moment is also defined as $\mu_{\alpha} = \sum_i q_i R_{i\alpha}$. We thus calculate $\epsilon^\infty_{\alpha\beta}$ and $Z^*_{i\alpha\beta}$ as single derivatives of the dipole calculated from a sum of partial charges,

\begin{equation}
    \epsilon^\infty_{\alpha\beta} = (\frac{d\mu_{\alpha}}{d\vec{E}_{el,\beta}} + I)\frac{1}{\epsilon_0},
\end{equation}

\begin{equation}
    Z^*_{i\alpha\beta} = \frac{d\mu_{\alpha}}{dR_{i,\beta}}.
\end{equation}
Thus the dielectric tensor and Born effective charges are accessible with the cost of atomic forces. We note that, due to the use of partial charges in the readout block of MPNICE, there can be numerical deviations between the dipoles computed via summation of partial charges versus via energetic derivative; in practice these are observed to be very slight.

\begin{figure}[!htb]
    \centering
    \includegraphics[width=0.75\textwidth]{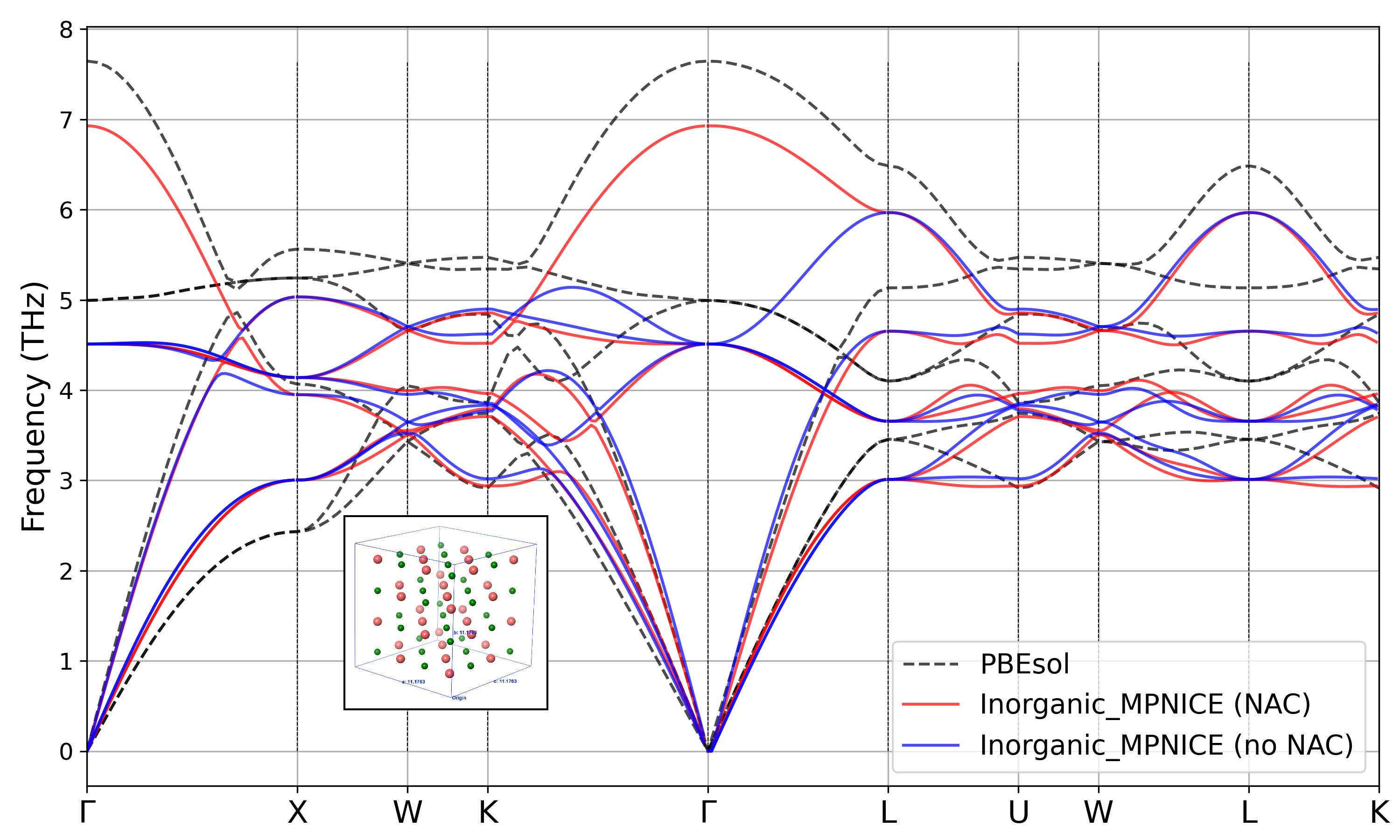}
    \caption{\MAT\ phonon band structure (THz) for NaCl (3x3x3 supercell pictured in inset), calculated using \MAT\ with and without the non-analytic correction (NAC) and compared to PBEsol calculated by ref. \citenum{petretto2018high}. Including the non-analytic corrections is necessary to obtain any LO-TO splitting at the gamma point. While the maximum phonon frequency from \MAT\ is slightly under-predicted versus PBEsol, most qualitative features are still present, and the LO-TO splitting is predicted within 0.24 THz.}
    \label{fig:NaCl_NAC}
\end{figure}

Although we do not train to any data with external electric fields, training \MAT\ to predict GFN1-xTB partial charges via the approximated Qeq method is sufficient to obtain qualitative response tensors. As an example, we compute Born effective charges and the fixed ion dielectric tensor for NaCl, using them to compute NACs to the phonon dispersion band structure (Fig. \ref{fig:NaCl_NAC}). Due to the use of a point charge representation, which cannot represent polarization orthogonal to interatomic vectors, \MAT\ exhibits small nonphysical off diagonal terms in the dielectric tensor. This effect is most extreme in systems with small symmetric unit cells, as in the case for NaCl. Additionally, likely due to being trained on xTB partial charges, the magnitude of $\epsilon^\infty_{\alpha\beta}$ and $Z^*_{i\alpha\beta}$ are reduced relative to PBE$_{\text{sol}}$ reference values (section S9). 
However, as the NAC relies on the ratio between $Z^*_{i\alpha\beta}$ and $\epsilon^\infty_{\alpha\beta}$, \MAT\ benefits from error cancellation and can reproduce the LO-TO splitting of NaCl fairly well.

\subsubsection{Application Highlight: Diffusion of Li ion in amorphous LiAlO2 cathode coating}

Inorganic coatings on the surface of electrode materials are frequently used in Li-ion batteries to prevent their degradation and improve performance.\cite{kaur2022surface} The transport of Li ions through the coating is a critical metric affecting the battery cyclic performance. Li conducting oxide materials such as LiAlO$_2$ have been used as an ultrathin conformal coating on the surface of electrode to improve the electrochemical stability of LNMO based cathode materials.\cite{park2014ultrathin} Prior studies have used AIMD simulations to calculate Li ion diffusivity. Due to high computational cost, these simulations have been performed using small supercells (64 atoms) for a few tens of ps. Here we demonstrate the use of \MAT\ to simulate the structure of the amorphous phase of LiAlO$_2$ (a-LiAlO$_2$). We have also calculated the diffusivity of Li ions in the a-LiAlO$_2$ at a wide range of temperatures using large scale molecular dynamics simulations. 

The unit cell of the crystal structure of LiAlO$_2$ was obtained from Materials Project (mp-3427) and a 5x5x4 supercell (1600 atoms) was created for MD simulations. The supercell was equilibrated at 300 K and 1 atm pressure for 500 ps using the NPT ensemble. The calculated density of the crystal phase at 300 K (2.584 g/cm$^3$) compares well with the value reported in the literature (2.615 g/cm$^3$)\cite{ha2020synthesis} as shown in Table \ref{tab:LiAlO2 properties}. Subsequently, the amorphous phase was generated by slowly heating the crystal phase from 300 K to 2800 K, where the system was equilibrated for 500 ps using NPT ensemble at each of selected temperatures (T=600 K, 1000 K, 1500 K, 2000 K, 2500 K and 2800 K). The volume of the crystal phase linearly increases as a function of temperature as shown in Fig. S2(c) and the coefficient of volumetric thermal expansion is computed using the slope of the linear fit. The snapshot of the amorphous structure after the equilibration at 2800 K and the Li-O RDF are shown in Fig. S2 (a) and (b), respectively. 

\begin{table}[!htb]
    \footnotesize
    \centering
    \begin{tabular}{|p{7cm}|c|c|}
    \hline
Property	&	\MAT	&	Literature	\\
    \hline
Density (g/cm$^3$)	&	2.584	&	2.615 (Expt)\cite{ha2020synthesis}	\\
    \hline
Coefficient of thermal expansion x 10$^{-5}$ (K$^{-1}$)	&	5.36	&	N/A \\
    \hline
Activation Barrier (eV)	&	0.55	&	0.56 (Expt) 0.54 (AIMD)\cite{park2014ultrathin} \\
    \hline
    \end{tabular}
    \caption{Comparison of calculated density, coefficient of thermal expansion and Li ion diffusion barrier with literature for LiAlO2 cathode coating material.}
    \label{tab:LiAlO2 properties}
\end{table}

The diffusivity of Li ions is calculated by using the 2800 K amorphous structure  as the initial structure for further equilibration using the NPT ensemble for 500 ps each at the desired temperature values (i.e. 2500 K, 2200 K, 2000 K, 1800 K, 1500 K, 1200 K, 1000 K and 800 K). Subsequently, the production simulations were performed at each temperature using NVT ensemble for 1 ns to obtain the trajectory needed for Li ion diffusivity calculations. The diffusivity is obtained by calculating the slope of the mean squared displacement of Li ions as a function of time over the length of production run at each temperature. The calculated Li-ion diffusivity (log scale) as a function of inverse of the temperature is shown in Fig. S2(d) for a wide temperature range of 800 K-2500 K. The computed diffusivity values can be fitted to Arrhenius type exponential function to obtain the activation barrier for diffusion. The computed activation barrier is in excellent agreement with the experimental value and the AIMD computed values from the literature as shown in Table \ref{tab:LiAlO2 properties}.

\subsection{Hybrid Inorganic and Organic model}\label{sec:hybrid}
In an attempt to build an MPNICE model which obtains reasonable accuracy on both inorganic and organic systems, we train a multi-task model with two output heads, one corresponding to \MPTRJ (PBE) and the other to the Lifesciences dataset ($\omega$B97X-D3BJ/def2-TZVPD). We refer to these two output heads as \MTLI\ and \MTLO, respectively. As the performance of such a multi-task model is dependent on the choice of output head to use, we additionally train a hybrid model with a single output head. To avoid conflicts between absolute energy scales, we simply remove the $\omega$B97X-D3BJ energies from the loss function, fixing the energy scale to that of the PBE dataset and relying on the atomic forces and dipole moments from $\omega$B97X-D3BJ to obtain accurate PESs. We refer to this method of training as ``shared force" training, and refer to the single output head model as \UNI.

\begin{table}[!htb]
    \scriptsize
    \centering
    \begin{tabular}{|p{3cm}|p{3cm}|c|c|c|c|c|}
\hline
 Test Class &   	Metric &	Inorganic	&	Hybrid	&	Hybrid$\_$I	&	Hybrid$\_$O	&	Organic	\\
\hline
\hline
Genentech & E RMSD (kcal/mol)	&	2.09	&	0.71	&	2.12	&	\textbf{0.30}	&	\textit{0.24}	\\
\hline
TorsionTest2000 & E RMSD (kcal/mol)	&	2.43	&	0.86	&	1.69	&	\textbf{0.47}	&	\textit{0.41}	\\
\hline
\multirow{2}{*}{TorsionNet500}& E RMSD (LOT) (kcal/mol)	&	2.38	&	0.88	&	2.15	&	\textbf{0.44}	&	\textit{0.33}	\\
\cline{2-7}
& E RMSD (CC) (kcal/mol)	&	2.35	&	0.89	&	2.15	&	\textbf{0.51}	&	\textit{0.43}	\\
\hline
Tautobase & E MAE (kcal/mol)	&	6.16	&	3.36	&	3.02	&	\textbf{0.58}	&	\textit{0.49}	\\
\hline
X23b & E RMSD	(kcal/mol) &	15.8	&	\textbf{10.6}	&	156.3	&	21.8	&	\textit{1.79}	\\
\hline
CSP & mean E RMSD (kcal/mol)	&	2.36	&	\textbf{2.09}	&	2.79	&	3.25	&	\textit{1.22}	\\
\hline
\multirow{3}{*}{Solvent MD} & $\lvert{\rho}\rvert$ $\%$ Error	& 35.2	&	26.5	&	23.7	&	\textbf{18.4	}&	\textit{4.9}	\\	
\cline{2-7}
& $\#$ failed$^\dagger$ &	\textit{0}	&	7	&	12	&	\textbf{2}	&	\textit{0}	\\
\cline{2-7}
& Water $\rho$ $\%$ Error	&	-7.0	&	20.3	&	$> 25\%$	&	\textbf{-14.0	}&	\textit{-6.5}	\\
\hline
\hline											
\multirow{2}{*}{Monoelemental SP} & E MAE (eV/atom)	&	\textit{0.022}		&	0.039	&	\textbf{0.036} &	0.373	&	N/A	\\
\cline{2-7}
& Mean E R$^2$	&	\textit{0.92}	&	\textbf{0.88}	&	\textbf{0.88}	&	0.83	&	N/A	\\
\hline
\multirow{3}{*}{Monoelemental Opt.}& V MAE (A$^3$/atom)	&	\textit{2.39}	&	3.39	&	\textbf{2.5	}&	3.25	&	N/A	\\
\cline{2-7}
& E MAE (eV/atom)	&	\textit{0.076}	&	\textbf{0.091}	&	0.096	&	0.451	&	N/A	\\
\cline{2-7}
& Mean E R$^2$	&	\textit{0.74}	&	\textbf{0.70}	&	0.66	&	0.69	&	N/A	\\
\hline
QMOF & E MAE (eV/atom)	&	0.030	&	\textit{\textbf{0.029}}	&	0.077	&	0.121	&	N/A	\\
\hline
\multirow{7}{*}{Elastic Moduli} & V MAE (A$^3$/atom)	&	\textit{0.50}	&	\textbf{0.54}	&	0.55	&	0.76	&	N/A	\\
\cline{2-7}
& $K_{VRH}$ MAE	&	21	&	\textit{\textbf{19}}	&	\textit{\textbf{19}}	&	25	&	N/A	\\
\cline{2-7}
& $K_{VRH}$ R$^2$	&	\textit{0.84}	&	\textit{\textbf{0.84}}	&	0.83	&	0.78	&	N/A	\\
\cline{2-7}
& $\#$ $K_{VRH}$ outliers	&	\textit{2}	&	\textbf{4}	&	5	&	13	&	N/A	\\
\cline{2-7}
& $G_{VRH}$ MAE	&	\textit{19}	&	25	&	\textbf{24}	&	\textbf{24}	&	N/A	\\
\cline{2-7}
& $G_{VRH}$ R$^2$	&	\textit{0.52}	&	0.36	&	\textbf{0.37}	&	\textbf{0.37}	&	N/A	\\
\cline{2-7}
& $\#$ $G_{VRH}$ outliers	&	\textit{82}	&	96	&	\textbf{93}	&	102	&	N/A	\\
\hline
\multirow{7}{*}{MDR Phonons} & $\%$ failure	&	0.09	&	\textit{\textbf{0.01}}	&	0.07	&	\textit{\textbf{0.01}}	&	N/A	\\
\cline{2-7}
& E MAE (meV/atom)	&	\textit{39}	&	292	&	\textbf{290}	&	581	&	N/A	\\
\cline{2-7}
& V MAE (A$^3$/atom)	&	\textit{0.35}	&	\textbf{1.1}	&	\textbf{1.1}	&	\textbf{1.1}	&	N/A	\\
\cline{2-7}
& $\omega$ MAE (K)	&	\textit{55}	&	80	&	\textbf{67}	&	77	&	N/A	\\
\cline{2-7}
& S MAE (J/K/mol)	&	\textit{49}	&	61	&	59	&	\textbf{50}	&	N/A	\\
\cline{2-7}
& F MAE (kJ/mol)	&	\textit{19}	&	25	&	24	&	\textit{\textbf{19}}	&	N/A	\\
\cline{2-7}
& Cv MAE (J/K/mol)	&	\textit{12}	&	16	&	15	&	\textbf{13}	&	N/A	\\
\hline
    \end{tabular}
    \caption{\scriptsize Results for the hybrid MPNICE models tested for the suite of organic (section \ref{sec:organic}) and inorganic (section \ref{sec:inorganic}) focused tests, compared to \MAT\ and \ORG. MPNICE is omitted from the model names for brevity. \textit{Italicized} values correspond to the most accurate model, while \textbf{bolded} values correspond to the most accurate hybrid model. $\dagger$ Some hybrid models predict liquid solvent densities resulting in $\simeq 1000$ atom cubic boxes with length $<$20\AA, which is unsupported in Desmond.}
    \label{tab:hybrid_tests}
\end{table}

First, we evaluate the relative performance for domain-specific organic and inorganic tests from sections \ref{sec:inorganic} and \ref{sec:organic}, the results of which can be seen in Table \ref{tab:hybrid_tests}. Note that the inclusion of the D3 correction term was dependent on the presence of dispersion corrections in the reference values for each test, regardless of the training set for the particular model; for example, D3 corrections were included for all models for the organic rotamer tests, and omitted for all models for the monoelemental single point test. For the majority of the domain-specific tests, it is clear that including disparate datasets results in a degradation of performance with respect to class-specific models. For the organic rotamer tests, disregarding the delta leanred model, \ORG\ obtains the best RMSD, but is closely followed by the $\omega$B97X-D3BJ output head of the multi-task model, \MTLO. The shared parameters between this and the PBE output head, \MTLI, do not result in significant improvement on these tests over \MAT, while \UNI, being trained directly to reproduce $\omega$B97X forces, does significantly improve performance on these fine energetic tests over \MAT. Notably, this performance does not extend to tautomer energies, where only \MTLO\ obtains reasonable errors out of the hybrid models. Additionally, albeit improving on \MAT, none of the hybrid models obtained reasonable errors for molecular crystal tasks, and exhibit 20-30$\%$ mean absolute errors on liquid solvent densities, similar to the PBE-D3 output head of the \CORG\ model (27\%), suggesting the performance is at least partially due to the intrinsic error of PBE-D3. \MTLO, the only output head trained directly to $\omega$B97X-D3BJ energies, achieves the best density error among the hybrid models, albeit with a 14$\%$ error for water, for which \MAT\ performs well with an error comparable to \ORG.

For the inorganic tests, \MTLI\ and \UNI\ generally perform the best out of the hybrid models, and primarily achieve very similar errors, with the notable exception of the QMOF test for PBE-D3 energies of metal organic frameworks. \MTLI\ PBE-D3 total energies for QMOF are effectively contaminated by the shared parameters with the $\omega$B97X output head trained to total energies, while \UNI, with its absolute energy scale fixed to that of PBE, slightly improves on \MAT. \MTLO, directly trained to reproduce $\omega$B97X energies, obtains poor MAE vs PBE energies, e.g. in the monoelemental and QMOF energy tests. Interestingly, \MTLO,  while systematically worse than the two hybrid models directly trained to PBE, still performs well on tests dependent on relative energetics and not total energies, such as energy correlation R$^2$ and mechanical properties. A good example is the MDR phonon test, for which \MTLO\ obtains comparable thermal properties, even outperforming the other hybrid models on some metrics.

Overall neither multi-task learning nor the shared force training result in a single output head which can reproduce the accuracy of both domain-specific models. From the tests presented here, we expect \UNI\ and \MTLO\ to be the most useful as a universal model, with the choice between the two dependent on the relative importance of the relative energies for organic or inorganic systems. Notably, \UNI\ performs well at inorganic total and relative energy tests, in many cases exceeding \MTLI, while still obtaining RMSDs for organic rotamer scans of less than 1 kcal/mol.

\subsubsection{Application Highlight: OLED emitter optimization}
\begin{figure}[!htb]
    \centering
    \subfloat[]{\includegraphics[width=0.3\textwidth]{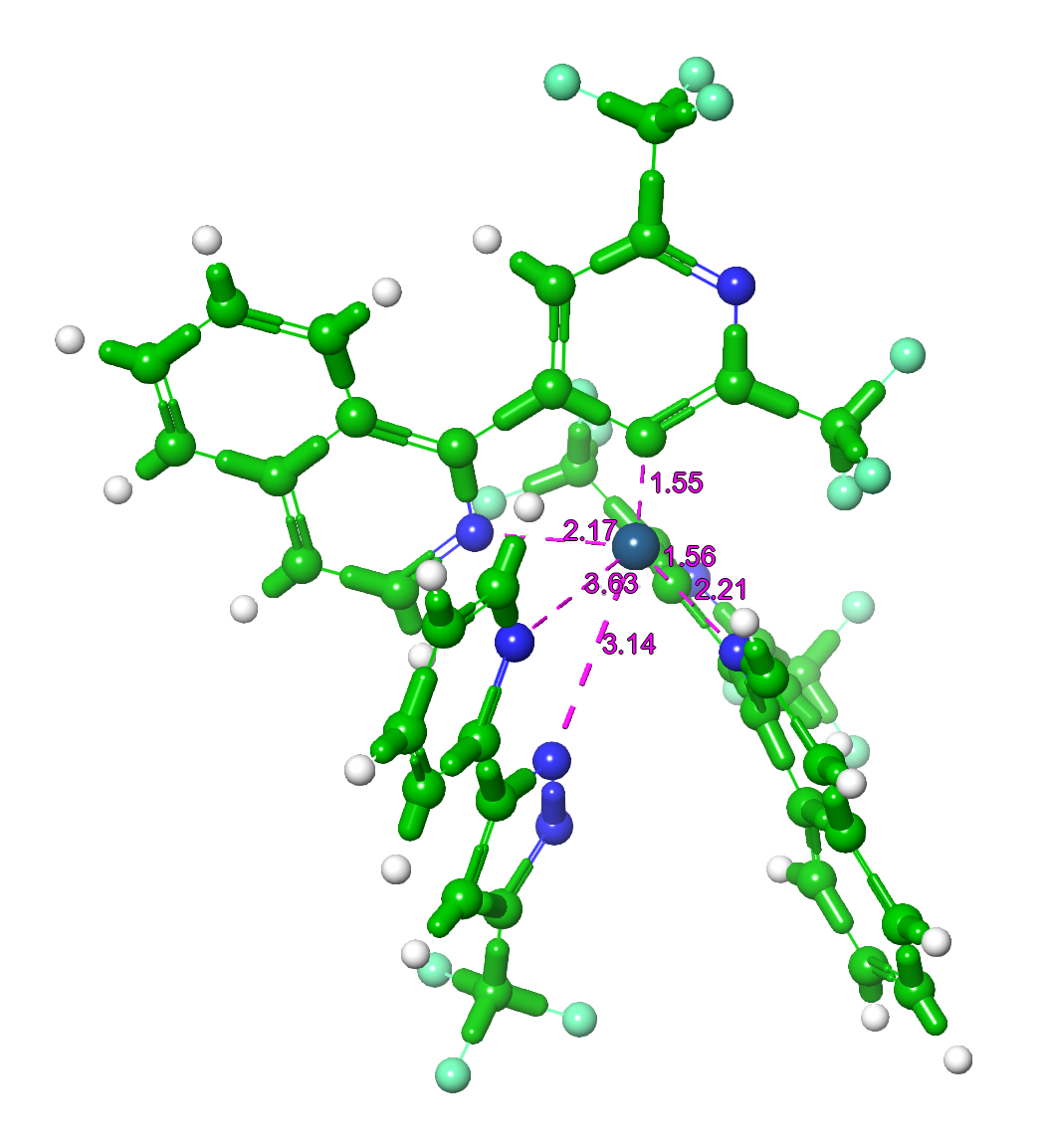}}
    \subfloat[]{\includegraphics[width=0.3\textwidth]{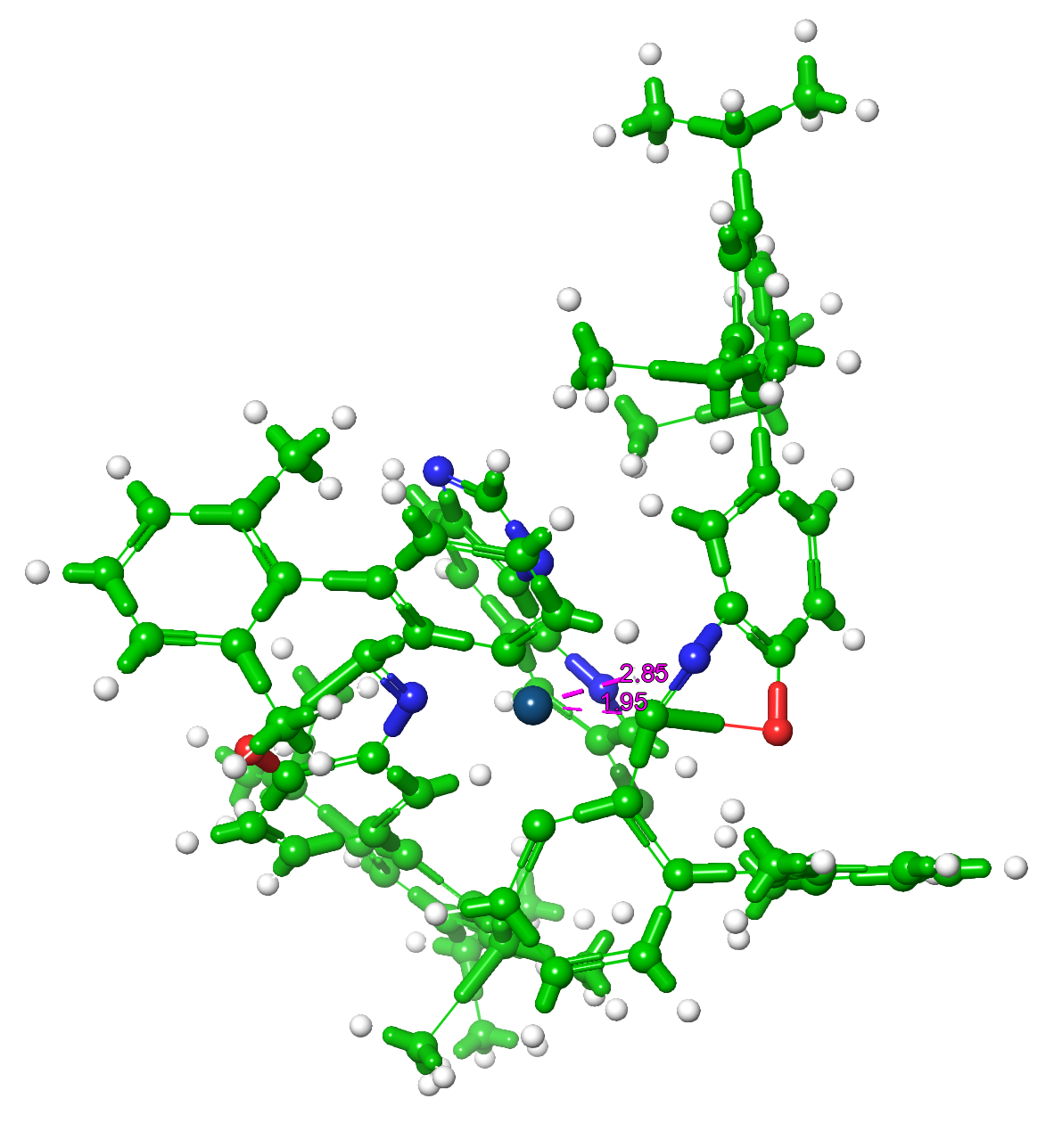}}
    \subfloat[]{\includegraphics[width=0.3\textwidth]{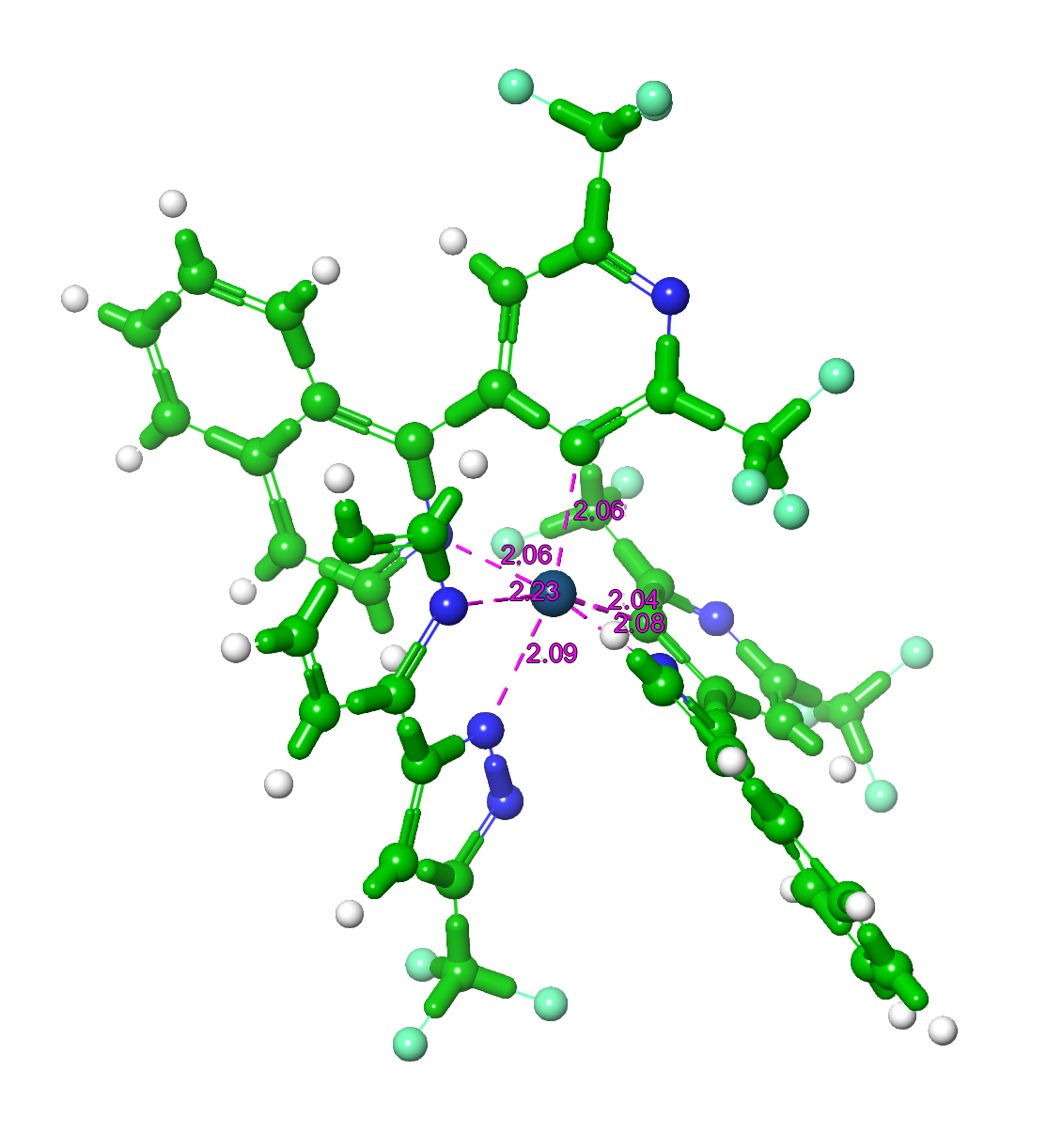}}
    \caption{\textbf{(a)} An optimized organometallic structure using \ORG. Metallic bonding was not present in the training set and so the ligand field is significantly distorted \textbf{(b)} an example structure with high RMSD optimized using \MTLI. The model incorrectly associates a carbon atom with Pt, at the expense of distorting the surrounding ring.  \textbf{(c)} an example structure optimized using \UNI. The ligand field is qualitatively correct without significant distortion of the ligands.}
    \label{fig:OLED}
\end{figure}

As the hybrid MPNICE models are trained on both organic and inorganic data, a natural test for generalization is to study systems which contain both. To this end, we consider a set of $\simeq500$ $\omega$B97X-D3/lacvp+$^{*}$ optimized unique gas phase organometallic molecules containing Pt(II) and Ir(II) centers, of relevance as emitters for OLED devices.\cite{damm2024opls5} We test each MPNICE model by optimizing these structures and calculating the average structural RMSD versus the DFT reference geometry. The results can be seen in Table \ref{tab:OLED}, with representative structures for \ORG\ and \MTLO\ in Figure \ref{fig:OLED}. Unsurprisingly, \MPT\ and \ORG\ obtain significantly worse RMSDs versus \UNI. \ORG, not having been trained on any data for Pt or Ir, does not obtain qualitatively correct metal-ligand bonding, with many complexes partially dissociating one or more ligands (Figure \ref{fig:OLED}a). Interestingly, the PBE output head of the multi-task model, \MTLI, gets significantly worse RMSD versus \MPT. This is observed to correspond to distorted ligand structures in the immediate vicinity of the metal (Figure \ref{fig:OLED}b). \MTLO, as well as \UNI, predict qualitatively correct metal-ligand bonding without distorting the organic ligands, suggesting that although the training sets do not explicitly contain organometallic complexes, they contain enough information about metallic bonding and organic molecules to provide qualitatively correct simulations. We expect \MTLO\ and \UNI\ to have sufficient accuracy to be useful, e.g. in pre-optimizing structures to reduce cost.

\begin{table}[!htb]
    \footnotesize
    \centering
    \begin{tabular}{|c|c|}
    \hline
Model	&	RMSD (\AA)	\\
    \hline
\MPT	& 0.301 \\
    \hline
\MTLI	& 0.509 \\
    \hline
\UNI	& 0.154 \\
    \hline
\MTLO	& 0.159 \\
    \hline
\ORG	& 0.324 \\
    \hline
    \end{tabular}
    \caption{Mean structural RMSD for a set of $\simeq500$ Pt(II) and Ir(II) organometallic complexes. Even though there is no organometallic data in the training sets, both \UNI\ and \MTLO\ obtain RMSD below 0.2\AA.}
    \label{tab:OLED}
\end{table}

\subsection{Inference Performance}\label{sec:profiling}

\begin{figure}[H]
    \centering
    \includegraphics[width=\linewidth]{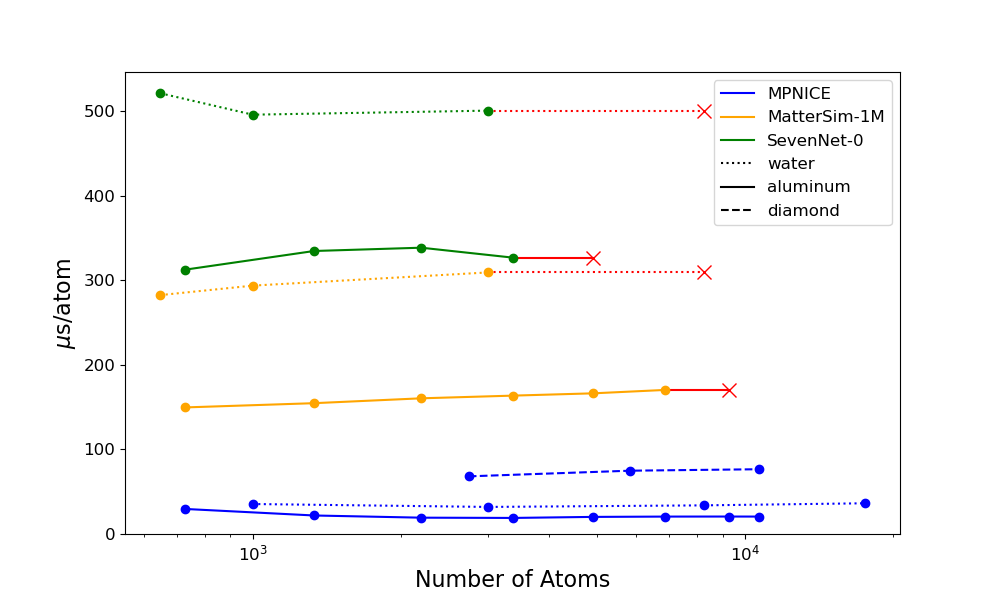}
    \caption{ Inference efficiency comparison of various MLFF models. A red `X' denotes a simulation which ran out of memory on an L4 GPU. \ORG\ was run for water, while \UNI\ was run for aluminum and diamond.}
    \label{fig:runtime_comparison}
\end{figure}

To evaluate the computational performance of MPNICE, we run a series of molecular dynamics simulations on an L4 GPU with 24 GB of VRAM, to represent performance on a typical mid-level card. Simulations were run for three prototype systems of different number density in order to profile the dependence on the number of nearest neighbors; water, diamond, and aluminum, similar to in ref. \citenum{mazitov2025pet}. Boxes were made to match experimental density, and each simulation was run using the NVT ensemble to maintain consistent number densities, and averaged over 500 steps. MPNICE is interfaced with the Desmond MD engine. For comparison, we also ran MatterSim-v1.0.0-1M and SevenNet-0, using their respective Atomic Simulation Environment (ASE)\cite{ase-paper} calculators. We chose these models due to their accessibility, comparable performance on inorganic benchmarks, and their being the fastest pretrained models currently available for both architectures. Inference times in $\mu$s/atom for boxes of increasing size can be seen in figure \ref{fig:runtime_comparison}.  Both MatterSim-v1.0.0-1M and SevenNet-0 run out of memory on the smallest diamond cell run with 2744 atoms. The Desmond implementation of MPNICE is seen to have competitive speed, asymptotically reaching 76, 36, and 20 $\mu$s/atom per timestep for diamond, water, and aluminum, respectively. The largest simulation run was for water, with 17,496 atoms; assuming a timestep of 0.5 fs this corresponds to a throughput of 0.07 ns/day using MPNICE. The smallest simulation was for aluminum with 729 atoms; assuming a slightly larger timestep of 2 fs, this corresponds to a throughput of 8 ns/day for MPNICE.

\section{Conclusions and Future Developments}\label{sec:conclusion}

MPNICE is an MLFF architecture developed to prioritize inference speed while simultaneously treating atomic charge as a first-class property and enabling general models trained to an arbitrary number of chemical elements. We have trained and benchmarked a suite of MPNICE models, with targeted models in each domain obtaining or approaching best-in-class accuracy, and inference speed an order of magnitude faster than comparable models. Additionally, iterative charge equilibration enables the direct prediction of charge dependent properties, such as the non-analytic correction to the phonon band structure of NaCl and vertical ionization energies of small molecules. 

For use in predicting relative energetics of conformations of finite organic molecules, as well as isomerization and tautomerization, we suggest the use of \ORGTB, with \ORG\ a close second, which can be used in cases where efficiency is paramount, or when periodic simulations are required.  For molecular dynamics simulations of organic liquids, we suggest the use of \ORG. For simulations of molecular crystals, including energetic ranking of polymorphs for organic crystal structure prediction, we suggest the use of \CORG\ (PBE-D3). For bulk inorganic materials and inorganic surfaces, we suggest the use of \MAT. For approximation of the structure of organometallic complexes, we suggest using \UNI\ or \MTLO.

We have not extensively explored the limit of MPNICE model capacity;  it is therefore possible that training significantly larger models, as in the recent 150 million parameter UMA models,\cite{wood2025family} would result in less loss of accuracy when training hybrid models such as \UNI. However, doing so will necessarily increase inference cost, limiting large-scale MD simulations and thus the accessible condensed-phase properties. While a broadly trained MLFF can yield reasonable simulation results for, e.g., simple liquids (see Sec.~\ref{sec:organic_MD}), the diversity of applications for general pretrained MLFFs virtually guarantees the necessity of fine tuning, particularly in targeted discovery campaigns requiring highly accurate predictions in complex environments such as battery materials and biological molecules. For such use-cases the advantage of general pretrained models lies in the learned internal representations (as in section \ref{sec:IPEA}), which allows significantly more data efficient training and model stability as compared to training a model from scratch. The ease of fine tuning, and the efficiency of the resulting model, relies on the efficiency of the initial architecture and pretrained models. Rapid evolution of MLFF technology over the next 3-5 years will reveal what is necessary and sufficient for efficient, predictive simulations of complex biological and materials science systems. We believe that the MPNICE architecture and pretrained models represent a useful step in this direction. 

While the results presented here are notable, there are some areas which require further development. In particular, MPNICE cannot rigorously distinguish systems with distinct spin states, and so is unable to predict quantities such as singlet-triplet gaps or magnetic ordering in materials. While \UNI\ and \MTLO\ have been demonstrated to achieve reasonable approximations of organometallic complexes, they still exhibit decreased performance versus domain-specific models, in particular failing to obtain liquid solvent densities within 10$\%$ of experimental values. Use of the presented pretrained models to unseen classes of material - such as reactive datasets, surfaces, and defects - are extrapolative in nature and thus subject to proportional skepticism, and fine-tuning is likely to be necessary for greater accuracy in some applications.
We expect this to be at least partially mediated in the future through incorporation of the rapidly evolving availability of additional training data, such as the recently released OMol25 dataset, which was reported after this work was completed and includes extensive organometallic and reactive sampling at the $\omega$B97M-V level of theory.\cite{levine2025open} 
Interfaces between organic liquids and materials, for which the reference level of theory is not well defined, are additionally not tested here. This limits confidence in zero-shot simulations of such interfaces, which are of significant importance in materials science. There is still much room to develop techniques to train a general MLFF which can effectively produce a single PES for systems with disparate reference levels of theory.


\section{Author Contributions}
J.L.W. conceptualized the model, developed software, trained and validated the inorganic and hybrid models and authored the manuscript. R.D.G. analyzed results for the matbench discovery benchmark. G. A. performed a case study on LiAlO$_2$.  Y.W. performed fine tuning for EA prediction and performed DLPNO-CCSD(T) computations on TorsionNet500.  A.A.F. developed software. X.X. curated water data, conceptualized the free energy algorithm and performed hydration free energy calculations.  J.S. developed software, curated data and edited the manuscript.  B.S. performed PBE-D3 computations on organic crystals.  L.D.J. conceptualized the project, developed software, curated organic data, trained and validated the organic models and authored the manuscript.  M.D.H. and R.A. reviewed and edited the manuscript.  L.D.J., R.A.F., M.D.H., K.L. and R.A. supervised the project.

\section{Conflicts of interest}
The authors declare the following competing financial
interest(s): R.A.F. has a significant financial stake in, is a
consultant for, and is on the Scientific Advisory Board of
Schrodinger, Inc.

\section{Code Availability}
\ORG, \ORGTB, \CORG, \MAT, \UNI, and both \MTLO\ and \MTLI\ are available for use in the Schrodinger Suite 25-2 release. 

\section{Data Availability}
Reference data for validation tests that are not already publicly available can be found at \href{https://github.com/leifjacobson/MLFF_test_data}{https://github.com/leifjacobson/MLFF\_test\_data}.  This includes reference energy data and geometries for the tautobase tautomer test, TorsionTest2000, DLPNO-CCSD(T) energies for TorsionNet500, the organic crystal test geometries for 20 crystal forming organic molecules with PBE-D3 energies as well as equilibrated initial geometries used for liquid density evaluation.
\begin{acknowledgement}
J.L.W. would like to thank Steven Dajnowicz for providing the geometries for Pt/Ir complexes.
\end{acknowledgement}

\section{Methods}
\subsection{Datasets}\label{sec:datasets}

\begin{table}[!htb]
    \footnotesize
    \centering
    \begin{tabular}{|c|c|c|c|}
    \hline
Datasets	&	Functional	&	Basis set	&	Properties	\\
    \hline
Lifesciences	&	$\omega$B97X-D3BJ	&	def2-TZVPD	&	E, F, $\mu$	\\
    \hline
Molecular Crystals	&	PBE-D3	&	plane wave	&	E, F, S	\\
    \hline
SPICE\cite{eastman2023spice,eastman2024nutmeg}	&	$\omega$B97M-D3BJ	&	def2-TZVPPD	&	E, F, $\mu$	\\
    \hline
OrbNet Denali\cite{christensen2021orbnet}	&	$\omega$B97X-D3	&	def2-TZVP	&	E	\\
    \hline
MP$_{trj}$\cite{deng2023chgnet}	&	PBE	&	See ref. \citenum{deng2023chgnet}	&	E, F, S, $q_{\text{GFN1-xTB}}$	\\
    \hline
OMAT24$_a$\cite{barroso2024open}	&	PBE	&	See ref. \citenum{barroso2024open}	&	E, F, S	\\
    \hline
    \end{tabular}
    \caption{Summary of datasets used to train in this work.}
    \label{tab:datasets}
\end{table}

A summary of the training sets used in this work is given in Table \ref{tab:datasets}. Two datasets, Lifesciences and Molecular Crystals, are new to this work; more details on their generation are given below. 

\subsubsection{Lifesciences Dataset}
We have initiated the dataset of non-periodic organic molecular systems (lifesciences training set) by relabeling portions of our prior lifesciences dataset,\cite{jacobson2022transferable, jacobson2023leveraging,zhou2025CSP} SPICE\cite{eastman2023spice,eastman2024nutmeg} and OrbNet Denali\cite{christensen2021orbnet} at a consistent level of theory (and also to add gradient data to the OrbNet Denali dataset) which we find is a good balance of accuracy and cost.  

The subsets of data which were relabeled were selected using data distillation, similar to what is described by Anstine \emph{et al.}\cite{anstine2024aimnet2}  Essentially, we train a model to some subset of a particular dataset and compute predictions on the full set.  Since the ground truth is known in the full set (albeit at a different level of theory than we plan to relabel at) we can compute the error on all examples and add any examples to the training set that have high error.  This procedure can be repeated until the error distributions on the training and test sets are sufficiently similar.

In addition to the above-described relabeling of datasets from three sources, we have performed active learning, utilizing preliminary MPNICE models, to improve the performance of the model in three distinct areas. First is to cover uncommon bonding patterns in molecules containing the atoms in the element set {H B C N O S F P Cl Br}, which are the most common elements appearing in drug-like organic molecules.  Second is to cover clusters of molecules representative of organic liquids and third is to cover interactions representative of molecular organic crystals.  Each of these subsets utilized a different active learning approach briefly described below.

In order to cover some bonding patterns not common to druglike datasets we have enhanced the ``dark matter universe" (DMU), developed along with the well known GDB-13 dataset.\cite{Fink2007GDB13}  We utilize substitution rules, similar to the construction of GDB-17\cite{Ruddigkeit2012GDB17} to expand the DMU to the element set stated above.  This source of molecules is very large and contains many examples of highly reactive molecules.  We selected all molecules with fewer than four non-hydrogen (heavy) atoms as well as 2500 additional, randomly selected molecules.  Each of these molecules was optimized and examples were selected from 20 ps of MD, resulting in 135,000 examples of small molecules.

To cover common intermolecular interactions of drug-like molecules we utilize a sampling protocol which involves performing single component molecular dynamics of disordered systems, followed by extraction of clusters.  Once a molecule is chosen we perform stereochemical enumeration and limited conformational sampling.  This yields a structurally diverse set of stereoisomers that we pack into a disordered systems (liquid).  This liquid is then optimized, followed by NVT MD at the optimized volume and 150 ps of NPT MD.  Clusters are extracted from the NPT dynamics.  Here, we select molecules in two ways.  We first filter our element enhanced DMU dataset, using filters similar to Fink \emph{et al.}\cite{Fink2007GDB13} to yield a set of relatively stable organic molecules.  We then select the following sets of molecules: all systems with three or fewer heavy atoms, a random selection in each of a number of special groups we found to be error prone with our prior MLFF (halogen containing molecules, halogenated aromatic groups, hydrazines, nitro groups, cyano groups and sulfonamides), and finally, a random sample of the full filtered, element-enhanced DMU.

Finally, we have included our dataset of electrolyte materials relevant to battery materials research in our lifescience set, as it was originally labeled at the same level of theory.\cite{dajnowicz2022high}.  All together, this yields about 11 million examples of finite systems labeled with energies, forces and dipole moments at the $\omega$B97X-D3BJ/def2-TZVPD level.

DFT calculations for finite systems labeled with $\omega$B97X-D3BJ/def2-TZVPD were performed with Psi4\cite{turney2012psi4} v1.6 using the same settings as in ref. \citenum{stevenson2022transfer}:
\begin{verbatim} `scf_type': `MEM_DF', `dft_basis_tolerance': 1e-10,
`ints_tolerance': 1e-10, `maxiter': 200, `dft_pruning_scheme': `robust',
`s_orthogonalization': `partialcholesky', `s_cholesky_tolerance': 1e-6
\end{verbatim}

\subsubsection{Molecular Crystals Dataset}
In order to improve the description of organic molecular crystals we have performed a search of high quality Z'=1 crystal structures from the CSD.  All structures with more than 200 heavy atoms in the unit cell are filtered.  The remaining crystals are relaxed and for about half of the systems we additionally ran short NPT MD simulations of the crystal and five frames of this dynamics are stored.  We compute energies, forces and stresses of the full crystals with PBE-D3 and also extract clusters for finite system labeling.  This procedure resulted in a dataset of about 430,000 organic crystal structures with periodic PBE-D3 training labels.  We have recently reported a procedure to perform crystal structure prediction on a large number of systems, with reference experimental structures largely extracted from the CSD.\cite{zhou2025CSP}  In order to keep some of this data as a true test, the CSD refcodes of the crystals studied in that work were removed from the set of structures used for active learning.  Additionally, no decoy structures from our prior work are included as training data.  As such, all examples in the periodic organic crystal training set are near experimental forms.

For the Molecular Crystal dataset we perform PBE-D3 energy, force and stress evalution with Quantum Espresso (QE). Ultrasoft pseudopotentials (GBRV type) are used with a wavefunction kinetic energy cutoff of 50 Ry and a k-point sampling density of 0.05 \AA$^{-1}$.  These settings are identical to those used in our CSP work.\cite{zhou2025CSP}

\subsection{Model training}\label{sec:training}
MPNICE models were trained using a multi-task Huber loss function with adaptive task weights,

\begin{equation}
\begin{aligned}
\mathcal{L} = \sum_{x} \Big[ &\frac{1}{N_{b}^x}\sum^{N_b^x}_i\frac{\mathcal{L}_{Huber}\left(\frac{\Delta E_i^x}{N_{a,i}^x},\delta_E\right)}{2\sigma_{E_x}^2} + 
\frac{1}{3\sum^{N_b^x}_iN_{a,i}^x}\sum^{N_b^x}_i\sum^{N_{a,i}^x}_j\sum^{\{x,y,z\}}_{\alpha}\frac{\mathcal{L}_{Huber}\left(\frac{\Delta F_{ij\alpha}^x}{|F_{ij\alpha}^x|},\delta_F\right)}{2\sigma_{F_x}^2} \\
&+ \frac{1}{9N_{b}^x}\sum^{N_b^x}_i\sum^{\{x,y,z\}}_{\alpha}\sum^{\{x,y,z\}}_{\beta}\frac{\mathcal{L}_{Huber}\left(\Delta S^x_{\alpha\beta},\delta_S\right)}{2\sigma_{S_x}^2}\Big] \\
&+ \frac{1}{\sum^{N_b^x}_iN_{a,i}^x}\sum^{N_b^x}_i\sum^{N_{a,i}^x}_j\frac{\mathcal{L}_{Huber}\left(\Delta q_{ij},\delta_q\right)}{2\sigma_{q}^2} +  \frac{1}{3N_{b}^x}\sum^{N_b^x}_i\sum^{\{x,y,z\}}_{\alpha}\frac{\mathcal{L}_{Huber}\left(\Delta \mu_{i\alpha},\delta_{\mu}\right)}{2\sigma_{\mu}^2} \\
&+\log(\sigma_{\mu}\sigma_{q}\prod_x\sigma_{E_x}\sigma_{F_x}\sigma_{S_x}),
\end{aligned}
\end{equation}\label{eq:loss}
where $x$ refers to a specific dataset, $N_b^x$ is the number of molecules in a batch belonging to the dataset $x$, $N_{a,i}^x$ is the number of atoms in molecule $i$ belonging to the dataset $x$, $\sigma$ are trainable parameters. Depending on the availability in the dataset, energies (E), forces (F), stress (S), atomic partial charges (q) and system dipole ($\mu$) are included in the loss function. In the case that neither partial charges nor system dipoles are available for any dataset being trained to, (e.g. for \OMA), the predicted charges are still trained via incorporation into the energy loss term, although in practice we have found that this can lead to flipped sign conventions. Following ref. \citenum{batatia2023foundation}, we set the Huber delta to 10 meV, with the force delta scaling down as the magnitude of the force in the system increases. Additionally, the loss function for the force is normalized by the absolute value of the reference force, minimizing the impact of large outliers in the training dataset. Datasets were split into training and validation sets using a random split with a 99:1 ratio. For organic models, the energy loss terms are not normalized to the number of atoms in the system; i.e. $\mathcal{L}_{Huber}\left(\Delta E_i^x,\delta_E\right)$ is used instead.

To minimize the loss function we use the AdamaxW optimizer (Adamax with decoupled weight decay) with a weight decay of 1.0e-4. We do not optimize batches to balance the frequency of seeing the various labels. We use exponential-moving-averaged weights and biases to evaluate validation and test errors, which are updated every 10 batches with a smoothing factor of 0.001. As is standard practice, we reduce the range of the energy labels and center them towards zero by subtracting an offset energy for each atom, defined by fitting a linear model for each level of theory.

\begin{table}[!htb]
    \footnotesize
    \centering
    \begin{tabular}{|c|c|c|c|c|p{3.5cm}|}
    \hline
Model	&	$N_{\mathrm{int}}$	&	$N_k$	&	$N_{\tilde{k}}$	&	MLP dimension	&	Dataset(s) trained to	\\
    \hline
\ORG	&	3	&	64	&	32	&	256, 192, 128, 112, 96	&	Lifesciences, Spice, OrbNet Denali	\\
    \hline
\CORG	&	3	&	64	&	32	&	256, 192, 128, 112, 96	&	Lifesciences, Spice, OrbNet Denali, Molecular Crystals	\\
    \hline
\ORGTB	&	3	&	64	&	32	&	256, 192, 128, 112, 96	&	Lifesciences, Spice, OrbNet Denali	\\
    \hline
\MPT	&	3	&	64	&	32	&	256, 192, 128, 112, 96	&	MP$_{trj}$	\\
    \hline
\OMA	&	3	&	128	&	32	&	288, 224, 192, 144, 112, 96	&	OMAT24$_a$	\\
    \hline
\MAT	&	3	&	128	&	32	&	288, 224, 192, 144, 112, 96	&	OMAT24$_a$ $\rightarrow$ MP$_{trj}$	\\
    \hline
\UNI	&	3	&	64	&	32	&	256, 192, 128, 112, 96	&	Lifesciences$^*$, MP$_{trj}$	\\
    \hline
\MTLO	&	3	&	64	&	32	&	256, 192, 128, 112, 96	&	Lifesciences, MP$_{trj}$	\\
    \hline
\MTLI	&	3	&	64	&	32	&	256, 192, 128, 112, 96	&	Lifesciences, MP$_{trj}$	\\
    \hline
    \end{tabular}
    \caption{Summary of models trained in this work. $^*$ denotes omission of energies from that dataset, whereas $\rightarrow$ denotes sequential training. $N_{\mathrm{int}}$ denotes the number of interaction blocks, $N_k$ is the number of node features, and $N_{\tilde{k}}$ is the number of features used in the messages, as in equations 5-6.}
    \label{tab:models}
\end{table}


\begin{suppinfo}

Materials project IDs for systems tested in the monoelemental structure ranking test. Details for the Torsion2000 set. Details for tautomer test. X23b cohesive energies. Details for the organic crystal tests. MPNICE densities for the set of 62 organic solvent simulations. Details on running DLPNO-CCSD(T) for TorsionNet500. Details for calculating the hydration free energy of water. Electric response tensors for NaCl using \MAT. Details on running the MDR phonon benchmark. Data for LiAlO$_2$ simulations. 
\end{suppinfo}

\bibliography{main}

\end{document}